# On the influence of several factors on pathway enrichment analysis

Sarah Mubeen[1,2,3,*], Alpha Tom Kodamullil[1], Martin Hofmann-Apitius[1,2], and Daniel Domingo-Fernández[1,3,4,*]

1. Department of Bioinformatics, Fraunhofer Institute for Algorithms and Scientific Computing, Sankt Augustin 53757, Germany
2. Bonn-Aachen International Center for Information Technology (B-IT), University of Bonn, 53115 Bonn, Germany
3. Fraunhofer Center for Machine Learning, Germany
4. Enveda Biosciences, Boulder, CO, 80301, USA

***Corresponding Author**: Mubeen, S. & Domingo-Fernández, D. Department of Bioinformatics, Fraunhofer Institute for Algorithms and Scientific Computing (SCAI), Sankt Augustin 53757, Germany. Email: sarah.mubeen@scai.fraunhofer.de and daniel.domingo.fernandez@scai.fraunhofer.de

**Keywords:** pathway enrichment, gene set analysis, pathway database, *omics* data, benchmark, gene set collection

## Abstract

Pathway enrichment analysis has become a widely used knowledge-based approach for the interpretation of biomedical data. Its popularity has led to an explosion of both enrichment methods and pathway databases. While the elegance of pathway enrichment lies in its simplicity, multiple factors can impact the results of such an analysis which may not be accounted for. Researchers may fail to give influential aspects their due, resorting instead to popular methods and gene set collections, or default settings. Despite ongoing efforts to establish set guidelines, meaningful results are still hampered by a lack of consensus or gold standards around how enrichment analysis should be conducted. Nonetheless, such concerns have prompted a series of benchmark studies specifically focused on evaluating the influence of various factors on pathway enrichment results. In this review, we organize and summarize the findings of these benchmarks to provide a comprehensive overview on the influence of these factors. Our work covers a broad spectrum of factors, spanning from methodological assumptions to those related to prior biological knowledge, such as pathway definitions and database choice. In doing so, we aim to shed light on how these aspects can lead to insignificant, uninteresting, or even contradictory results. Finally, we conclude the review by proposing future benchmarks as well as solutions to overcome some of the challenges which originate from the outlined factors.

# 1. Introduction

Pathway enrichment analysis has become one of the foremost methods for the interpretation of biological data as it facilitates the reduction of high-dimensional information to just a handful of biological processes underlying specific phenotypes. Over the last decade, the popularity of pathway enrichment analysis has led to the development of numerous different methods that can be categorized into three generations: i) over-representation analysis (ORA), ii) functional class scoring (FCS) and iii) pathway topology (PT)-based [1]. Each of these generations add an increasing layer of complexity to the analysis, with ORA approaches often being 2x2 table methods, FCS methods including those which rely upon the coordinated activity of genes in a gene set, and PT-based, characterized as methods which take into account pathway topology. While the simplicity and accessibility of enrichment methods have been the main drivers to their widespread adoption by the community, the broad pool of methods at hand and the lack of gold standards pose a challenge in evaluating the variability of enrichment results. Consequently, several guidelines have been published in recent years on recommendations for the experimental design of an enrichment analysis [2-4].

An analogous but more philosophical debate in the community pertains to the choice of pathway or gene set database. Its selection is arguably one of the most decisive factors influencing the results of enrichment analyses as it determines the possible gene sets that can be enriched (i.e., genes within a gene set are enriched in an examined list of genes). The number of public databases has continued to grow in the past years in parallel with novel enrichment methods. However, the list of the most widely used databases has not changed in the last decade as enrichment analyses are predominantly conducted exclusively on one of the following three databases: KEGG [5], Reactome [6], and Gene Ontology (GO) [7]. While this selected group of databases come with several advantages (e.g., large coverage of biological processes and regular updates), definitions of what constitutes a given pathway or gene set may be arbitrarily drawn across databases.

At present, users are offered a wide spectrum of enrichment methods and databases when performing enrichment analyses. This poses a challenge when considering the numerous factors that play a role in results of enrichment analysis which can lead to insignificant, irrelevant or even contradictory results. Thus, in recent years, several benchmarks studies have been conducted to evaluate the effects of various aspects of pathway analysis for practical guidelines.

In this work, we review the findings of major benchmarks conducted on different factors that influence the results of pathway enrichment analysis (**Figure 1**). The goal of our paper is to both inform the broader community of researchers using pathway enrichment analysis of these factors as well as to summarize the findings of all the most recent benchmarks. Finally, we also discuss possible solutions to address these factors as well as other factors that have not yet been investigated but can be benchmarked in the future.

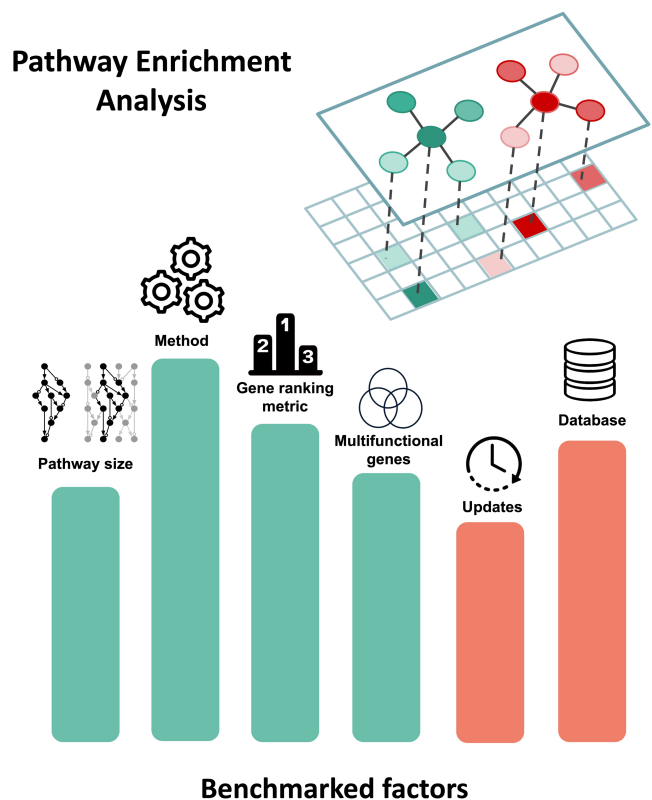

Figure 1. Illustration of major factors that influence the results of pathway enrichment analysis discussed in this review.

## 2. Comparative studies on enrichment methods

Given the popularity of pathway enrichment analysis, at least 70 different methods have been developed as well as hundreds of variants [8-9] (see Xie *et al*. [10] for an exhaustive survey of methods, tools, platforms and benchmarks). The implementations of these methods can differ based on a number of factors, such as the gene-level statistic (e.g., T-test statistic and fold change), the gene set-level statistic (e.g., Kolmogorov-Smirnov (KS) statistic [11] and Wilcoxon rank sum test [12]), the formulation of the null and alternative hypotheses, the significance estimate and how multiple testing corrections are applied. Many of the most commonly employed pathway enrichment methods have been compared in several major benchmarks and reviews. In this section, we outline the findings of 12 comprehensive comparative studies on enrichment methods **(Table 1; see Supplementary Table 2 for more details)**.

| 3. | Review | Methods tested | Datasets | Database (# pathways) |
|---|---|---|---|---|
| 1 | [13] | 7 | 36 | KEGG (116) |
| 2 | [2] | 10 | 75 | KEGG (323) and GO (4,631) |
| 3 | [3] | 7 | 118 | KEGG (232) |
| 4 | [14] | 6 | 10 | KEGG (86) |
| 5 | [15] | 9 | 3 | KEGG (114) |
| 6 | [16] | 13 | 6 | GO gene set collection, extracted from MSigDB [17]; v6.1) (5,917) |

| 7 | [18] | 8 | 3 | MSigDB v5.0 (10,295) |
| 8 | [9] | 10 | 86 | KEGG; 150 pathways for all methods except 130 for PathNet [19] and 186 for CePa [20-21] |
| 9 | [22] | 11 | 1 | C2 collection from MSigDB v4.0 (4,722) |
| 10 | [23] | 16 | 42 | KEGG (259) and Metacore$^{TM}$ (88) |
| 11 | [24] | 5 | 6 | KEGG (192) |
| 12 | [25] | 7 | 38 | KEGG (189) |

**Table 1. Comparative studies evaluating differences across enrichment methods**. In the third column, we report the number of enrichment methods compared in each study (see **Supplementary Table 2** for details on the specific methods tested in each study). Here, we would like to note that we differentiate between methods and tools/web applications based on Geistlinger *et al.* [2]. In the fourth column, we report the number of datasets each study performed comparisons on, all of which were real datasets except in Ihnatova *et al.* [3] and Rahmatallah *et al.* [22] which included both real and simulated datasets. Finally, the fifth column reports the pathway databases used in each study while the number of pathways is shown between parentheses.

## 3.1. Metrics for method evaluation

A particular challenge in the design of comparative studies on enrichment methods is that in the absence of a comprehensive understanding of the complex biological processes involved in a given phenotype, results are often not verifiable beyond retrospective evaluations. That is to say, without a gold standard with which to compare the results produced by any given method, conclusive assessments are often difficult to make. Nonetheless, several techniques to benchmark methods are widely used; in line with Tarca *et al.* [23], the majority of studies evaluate the performance of an enrichment method based on at least one of the following metrics: prioritization, specificity or sensitivity.

Prioritization is evaluated based on whether a target gene set that has been identified *a priori* as showing high relevance to a phenotype associated with the dataset under investigation is ranked near the top (e.g., the breast cancer pathway is expected to hold the topmost ranking for a dataset measuring transcriptomic differences between breast cancer patients and healthy controls). Specificity refers to the proportion of gene sets that are correctly identified by a method as true negatives; thus, methods with a high specificity will generate fewer false positives. Finally, of all the gene sets detected as significant by a given method, sensitivity measures the proportion of gene sets that are actually relevant to the phenotype associated with the dataset under study (i.e., true positives).

Of the various comparative studies done to date, the above-mentioned metrics have been among the most commonly used for the empirical evaluation of enrichment methods. Nonetheless, the metrics used and the methods benchmarked by an individual study can vary greatly, with the most popular methods, not surprisingly, studied the most frequently. Yet despite the numerous benchmark studies conducted thus far, a comprehensive and standardized assessment of the many enrichment methods available has yet to be performed. Moreover, of the benchmark studies which have attempted such an assessment, no specific method has been shown to yield consistent results across all evaluated settings. Nevertheless, trends do emerge regarding the individual performance of a method on a given metric **(Supplementary Tables 3-5)**. Thus, in the following, we report the trends observed across comparative studies for methods that consistently show superior performance on metrics in two or more studies without showing a poor performance on that same metric.

With regard to sensitivity, MRGSE [26], GlobalTest [27] and PLAGE [28] ranked highly in studies by Tarca *et al.* [23] and Zyla *et al.* [25] **(Supplementary Table 3)**. However, high sensitivity may also imply a lower specificity. This was indeed observed for MRGSE and PLAGE, both of which reported a larger than expected number of false positives in at least one study, though also a good performance in prioritization **(Supplementary Table 5)**. This is not surprising given that both methods have also been shown to report a majority of gene sets as significant [24-25]. Similarly, classical statistical tests, including the KS test and the Wilcoxon rank sum test, were highly sensitive in Bayerlová *et al.* [13] and Nguyen *et al.* [9], though results were inconsistent regarding their specificity. In Rahmatallah *et al.* [22], KS also demonstrated relatively good sensitivity which tended to remain relatively stable despite a decreasing sample size, though the false positive rate (i.e., 1-specificity) increased. Notably, of the above-mentioned methods, GlobalTest was the only investigated method to consistently demonstrate high sensitivity as well as high specificity in studies by Tarca *et al.* [23] and Zyla *et al.* [25].

In assessments of specificity, SPIA [29] and CAMERA [30] have shown high specificity in at least two studies **(Supplementary Table 4)**, though results have been mixed or poor with regard to sensitivity and target pathway prioritization. Furthermore, GSA [31], PADOG [32] and PathNet showed good results with regard to prioritization **(Supplementary Table 5)**, but mixed results for sensitivity and specificity. Finally, across all studies, GSEA [33] and ORA (or a variant) were the most investigated enrichment methods, with 8 of 12 comparative studies assessing either one or both of these methods **(Supplementary Table 2)**. Here, we observed that, although they were the most commonly used methods for enrichment analysis, results regarding their sensitivity, specificity and prioritization were altogether inconsistent **(Supplementary Tables 3-5)**.

## 3.2. Hypothesis testing and significance assessment

Much of the focus of comparative analyses on gene set analysis methods has been on the implications of alternative definitions of the null hypothesis. In their seminal work, Goeman and Bühlmann [34] characterized methods by the null hypothesis assumed in the statistical test. Enrichment methods, they assert, can be categorized as being competitive methods if they test the competitive null hypothesis, (i.e., those which assume that genes in a gene set are not differentially expressed with respect to their complement (typically the rest of the genes in the experiment)), or self-contained methods if they test the self-contained null hypothesis (i.e., those which assume that genes in a gene set are not differentially expressed across phenotypes). Choosing one category of methods over another can confer several advantages, which we explicate through a brief review of studies that have assessed the performance of methods which differ based on this distinction.

Rahmatallah *et al.* [22] recapitulated earlier work [35-37], generally noting that the power of self-contained methods was greater than that of competitive ones **(Table 1; Supplementary Table 2)**. Self-contained methods were also more robust to sample size and heterogeneity, with these methods showing the highest sensitivity amongst all the ones they evaluated, even as the sample sizes decreased [22] **(Supplementary Table 6)**. Specifically, they found that ROAST [38] and SAM-GS [39] yielded the best performance on this metric.

Geistlinger *et al.* [2] noted that the proportions of gene sets reported as significant by methods differed based on the type of null hypothesis tested. Out of ten investigated methods **(Supplementary Table 2)**, they found that the majority of self-contained ones, including GlobalTest, detected a larger fraction of gene sets as significant. In Zyla *et al.* [25], the self-contained methods GlobalTest and PLAGE also reported the largest number of gene sets as significant amongst all benchmarked methods **(Supplementary Table 2)**. In contrast to these findings, Wu and Lin [37] found that GlobalTest reported fewer gene sets as significantly enriched in comparison with competitive methods.

Furthermore, Geistlinger *et al.* [2] found that self-contained methods, particularly GlobalTest and SAM-GS, were especially sensitive to gene set size, with a propensity towards detecting larger gene sets as significant **(Supplementary Table 7)**. For example, even when random gene sets were assembled, GlobalTest and SAM-GS identified all gene sets with over 50 genes as significant. However, Maleki *et al.* [16] noted that GlobalTest was among the methods more likely to identify gene sets of smaller sizes as significant **(Table 1; Supplementary Table 2)**, albeit, in this case, the upper bound for genes in a given gene set was nearly 2,000, while in Geistlinger *et al.* [2], it was 500.

These contradictory findings are a prime example of the challenges associated with benchmarking methods for gene set analysis. Such glaring variability in results yielded by the same method investigated in different studies may be due to several factors, such as gene set size or differing proportions of differentially expressed genes in the studied datasets. For instance, GlobalTest tends to perform sub-optimally when only a few genes in a given gene set are differentially expressed and the majority of genes are not and conversely tends to be better suited for when there are many genes with small changes in differential expression in a gene set [37, 40]. We further discuss the impact of gene set size on results irrespective of the null hypothesis tested in subsequent sections.

If opting to select a competitive method instead, one must consider that testing the competitive null hypothesis often inherently implies not only the intended association between the phenotype and the genes within a given gene set, but also between the phenotype and genes in the complement of the set [40]. That said, competitive methods can be appropriate when the goal is to test for excessive amounts of differential expression among genes in a gene set. For instance, the popular ORA method was noted as suitable when there are large levels of differential expression [2]. However, ORA also tends to prioritize larger gene sets, assigning them lower *p*-values [16, 23]. Nonetheless, in Geistlinger *et al.* [2], ORA and other competitive methods outperformed self-contained ones in ranking phenotype relevant gene sets near the top **(Supplementary Table 8)**. In contrast, although ORA performed favourably on the prioritization of relevant gene sets in Tarca *et al.* [23], no clear discernment could be made with regard to the performance of competitive and self-contained methods on this measure **(Supplementary Table 5)**. Furthermore, while self-contained methods tended to identify a larger proportion of gene sets as significant in Geistlinger *et al.* [2], the majority of competitive methods (i.e., SAFE [41], GSEA, GSA and PADOG) did not identify any significant gene sets.

Intimately linked to the formulation of the null hypothesis is the calculation of the *p*-value [34]. Divergent approaches to assign a *p*-value to a gene set address the following question: what is the sampling unit? If the sampling unit is the gene, for each gene set of a given size, an equal number of genes are randomly drawn from all genes under investigation to sample the null distribution. If however, the sampling unit is the subject, the phenotypic labels of subjects are randomly permuted to sample the null distribution instead. While methods which test a self-contained null hypothesis are generally linked with sample permutation and competitive methods with gene permutation, the latter group of methods can be modified to make them self-contained [40].

Sample permutation is often regarded as the preferred approach to obtain the empirical null distribution as its setup tends to pertain more naturally to the research question at hand of whether or not an association exists between a gene set and a phenotype. In contrast, methods which calculate significance by gene permutations suffer from the assumption that genes are independent and identically distributed (*iid*), often resulting in a large number of false positives [37, 40, 42-43]. It is well-established, however, that this premise does not hold true in a real biological context where genes tend to be correlated and which is indeed the intended goal of enrichment analysis of identifying sets of genes working in tandem [37]. Thus, in the case of gene permutations, while significant gene sets may be reflective of either gene correlations and/or actual phenotypic differences, it is the latter outcome which is often far more interesting.

The effects of correlations within gene sets have been observed in various studies. Tamayo and colleagues [44] show that these correlations can have major implications on the results of enrichment analysis by comparing the results of GSEA against a simple parametric approach in 50 datasets. They observed that the parametric approach, which assumes differential gene expression scores are both independent and follow a normal distribution, yields a larger number of significant pathways than GSEA, but many of these are speculated to be false positives. Similarly, in experiments on simulated data in Maciejewski [40], the author demonstrated that when gene correlations were present in the gene set yet there were no differentially expressed genes either in the gene set or its complement, false positive rates for methods which make the *iid* assumption (e.g., parametric methods proposed in Irizarry *et al.* [45] and competitive methods with gene permutation) were greater than expected. Thus, the authors of these studies caution that methods that assume gene independence may report gene sets as significantly associated with a phenotype when in fact gene correlations account for the purported, significant results. However, it is also worth noting that the influence of correlations can be somewhat mitigated by reducing redundancies within gene sets.

In Maciejewski [40], the author observed that among methods with a sample permutation procedure, GlobalTest, GSEA, and GSA and its variant achieved higher power. Furthermore, GSEA, a competitive method with sample permutation, had higher power than several other methods tested (i.e., GSA and its variant, PAGE [46], Wilcoxon rank sum test, Q1 [47], and SAFE), although as the number of differentially expressed genes in a gene set increased, so too did the power of the other methods.

Nevertheless, sample permutation requires an adequate number of samples as without a sufficiently large sample size, the calculated *p*-value may never achieve significance, in which case, gene permutation is recommended. For instance, in their comparative analysis, Maleki *et al.* [48] found that, across 10 replicate datasets, GSEA with sample permutation was unable to detect any gene set as enriched when sample sizes were small, suggesting a lower bound of 10 samples for this particular method. The robustness of various methods to changes in sample size is further discussed in subsequent sections.

Other methods have been proposed that attempt to address some of the drawbacks associated with sample and gene permutation approaches by conducting both sample permutations and gene randomizations in a method known as restandardization, as with GSA, through the use of rotations for gene set testing, as with FRY [49] and ROAST, or via bootstrapping methods, as in Zahn *et al.* [50] and Barry *et al.* [43].

### 3.3. Topology and non-topology -based methods

Methods for enrichment analysis can also be classified as those which are topology-based or non-topology-based. The latter group of methods can be further sub-classified into the aforementioned ORA and FCS methods, the so-called first and second generation approaches, respectively [1]. PT or topology-based methods fall into the category of third generation approaches, intuitively more advanced as, unlike ORA and FCS methods, they leverage the topological structure of genes in a pathway. Nonetheless, results from multiple benchmarks on topology and non-topology -based methods are inconclusive as to the superiority of one group of methods over another, with studies suggesting topology-based methods have the upper hand.

In Bayerlová *et al.* [13], authors noted that whether a method was topology-based or not was inconsequential to performance when original KEGG pathways (which tend to contain overlapping genes) were used in experiments **(Supplementary Tables 2-5)**. Notably, while CePa includes pathways from both KEGG and the Pathway Interaction Database (PID), other topology-based methods evaluated in the study (i.e., PathNet and SPIA) are only compatible with pathways formatted in a custom-XML format (i.e., KEGG Markup Language). This result is particularly striking considering KEGG contains overlapping pathways, thus limiting the potential of topology-based methods by restricting users to pathways formatted in the manner specified by this database. In contrast, experiments done using non-overlapping pathways resulted in topology-based methods outperforming non-topology-based ones [13]. Inline with these findings, comparative studies by Jaakkola and Elo [14] and Nguyen *et al.* [9] similarly suggested topology-based methods exhibit an improved performance over non-topology-based ones under certain conditions, albeit, contrary to findings by Bayerlová *et al.* [13], these conclusions were drawn exclusively using KEGG as the choice of pathway database.

More particularly, results from Nguyen *et al.* [9] indicate that topology-based methods have a slight upper hand in detecting target pathways as compared to non-topology-based ones **(Supplementary Table 5)**, though results were mixed regarding the *p*-values of target pathways. In Jaakkola and Elo [14], topology-based methods (i.e., SPIA, CePa and NetGSA) detected a larger number of significant pathways than non-topology based ones (i.e., GSEA, Pathifier and DAVID). However, in a more challenging dataset where differences across groups were subtle, nearly all studied methods identified either no pathways or relatively few pathways as significantly enriched.

Ihnatova *et al.* [3] conducted several experiments which assessed the influence of various parameters on topology-based methods (e.g., sensitivity to pathway and sample size **(Supplementary Table 6)**), specificity **(Supplementary Table 4)**, and exclusion of topological information). As a proxy to study the latter parameter (i.e., whether topological information affects results for a given topological method), the authors evaluated the influence of single genes on the fraction of pathways that were considered enriched, assuming that a setup which fails to take into account pathway topology is one in which individual genes have an equal impact on results. To that end, they found that TopologyGSA [51] and Clipper [52] yielded no difference in performance when topological information was excluded, while for all other methods, the exclusion of topological information led to the identification of a smaller fraction of enriched pathways. Additionally, in assessing whether the ranks/*p*-values of target pathways change when topological information is incorporated, the authors found that both the ranks and *p*-values of target pathways decreased for PRS [53] and CePa, while for all other methods, the inclusion of topological information resulted in either no change or an increase in ranks/*p*-values of target pathways (at times caused by pathway-specific effects).

## 3.4. Additional methodological considerations for enrichment analysis

Besides the above-mentioned common measures and classifications several comparative studies have used to draw distinctions between enrichment methods, the performance of methods on a number of additional aspects have also been compared, which we describe in this section. Furthermore, we refer to the studies which evaluate other aspects in **Supplementary**

**Table 10**, including accuracy **(Supplementary Table 9)**, type I error rate, power, runtime, and assessments of reproducibility across datasets, amongst others.

**Sensitivity to sample size**

Several studies have specifically investigated the impact of sample size on method performance. Not surprisingly, performance tends to decay as sample sizes decrease, though some methods tend to be more robust than others with regard to *p*-values of target pathways [25], the true positive rate [22], and proportions of enriched pathways [3]. A survey of the performance of methods based on their robustness to sample size can be found in **Supplementary Table 6**. Here, again, we describe the trends noted across studies concerning the performance of methods (i.e., those which show good performance in at least two studies), finding MRGSE, ORA, GSEA with gene permutations (GSEA-G) and PADOG to be among the non-topology based methods with consistently good performance. The influence of sample size was also investigated specifically for topology-based methods in Ihnatova *et al.* (2018) with regard to the proportion of enriched pathways detected. Here, SPIA, PRS, and CePa were the most robust against this factor. By contrast, TopologyGSA, Clipper, DEGraph, and TAPPA were highly sensitive to increasing sample size [3].

In Maleki *et al.* [48], the authors note that the reproducibility of a method suffers when groups contain fewer than five samples. Furthermore, for some methods (i.e., GSVA [54], GAGE [55], FRY, and ROAST), they observed that an increase in sample size tends to be accompanied by an increase in the number of significant pathways, though this likely implies a greater number of false positives. In contrast, PLAGE and ssGSEA detected nearly all gene sets as enriched regardless of the sample size. As these latter two methods were found in studies by Tarca *et al.* [23] and Zyla *et al.* [25] to be highly sensitive **(Supplementary Table 3)**, these results may not be altogether surprising.

**Sensitivity to gene set size**

Many of the studies we review **(see Table 1)** have also compared the performance of enrichment methods by their sensitivity to gene set size **(Supplementary Table 7)**. Across studies, little consensus could be noted on this factor, complicated by variability in how set sizes are binned and definitions of what constitutes, for example, a large gene set. For instance, in Ihnatova *et al.* [3], large pathways in their studies were those that contained >= 35 genes (and up to 344), in Bayerlová *et al.* [13], that number was over 80 (and up to 380), whereas the sizes of gene sets in Geistlinger *et al.* [2] and Maleki *et al.* [16] reached 500 and nearly 2,000, respectively. Thus, objective comparisons can be difficult to make, and here instead, we briefly discuss study-specific trends related to this parameter.

In Maleki *et al.* [16], the authors observed that some methods tended to preferentially report gene sets of larger sizes as significant (i.e., FRY, ROAST, GlobalTest and GSEA-G), while others tended to report smaller ones (i.e., GAGE and ORA). Though it may be the case that a method shows poor robustness against this parameter, a method itself may still be well-suited for gene sets within a specific range of sizes. For example, although ORA performed suboptimally for smaller gene sets in Tarca *et al.* [23], both ORA and GSEA-G ranked as the top methods when relevant gene sets were larger in size. Additionally, studies by Ihnatova *et al.* [3] and Ma *et al.* [15] compared topology-based methods on this metric. Overall, larger pathways were found to have smaller *p*-values for all topology-based methods investigated, except PRS in Ihnatova *et al.* [3], while Ma *et al.* [15] assessed their performance with respect to gene expression as well as metabolomics/lipidomics data, a comparison closely related to pathway size (i.e., focus is on smaller pathways in the latter case). The authors found that all investigated methods generally tended to perform about the same in the case of large pathways and their previously published method, NetGSA, as well as DEGraph, held an upper hand over other methods in the case of small biochemical pathways. Due to the substantial implications of gene set size on the results of enrichment analysis, this factor is further discussed in a subsequent section.

**Data preprocessing**

Another important consideration in performing enrichment analysis is in the selection of the steps used in data preprocessing. For microarray data, these steps include background correction to remove noise, and normalization to reduce variations from biological and technical factors that can occur in the measured intensities of gene expression [56]. Methods used for major preprocessing steps include RMA [57], its variant, gcRMA [58], MAS 5.0 (Affymetrix; version 5), and variance stabilizing normalization (VSN) [59] [60]. With regard to the preprocessing of RNA-seq data, several different approaches have been benchmarked in a comprehensive review by [61]. Here, the authors contend that the normalization method selected can substantially impact downstream analyses, finding that DESeq [62] and TMM [63] generally display good performance across a variety of measures whilst relatively poor performances were noted for RPKM [64] and Total Count [61].

**Applicability of various *omics* dataset types to enrichment analysis**

Given the steady shift from hybridization approaches to RNA sequencing, the applicability of enrichment methods, initially developed for microarrays, to RNA-seq data was specifically studied in Rahmatallah *et al.* [22]. The authors found that, with appropriate normalization, RNA-seq data is just as suitable as microarray data as input for these methods. It is worth noting that while the studies we have reviewed have largely been performed on real and/or simulated RNA-seq and microarray data, more recent studies have analyzed the performance of enrichment methods on other data types. For instance, a study by Holland *et al.* [65] ventures into an assessment of the applicability and performance of enrichment analysis on single cell RNA-seq data. Much in line with other benchmarks, they find that the gene sets selected are more sensitive to the analysis than the underlying statistic. Additionally, in Mora [66], the author also reviewed the use of methods for enrichment analysis besides RNA-seq and microarrays, including ChIP-Seq, SNP, methylation and non-coding RNA data, while Maksimovic *et al.* [67] benchmark the performance of enrichment analysis methods specifically designed for methylation data.

**Gene – level statistics**

Depending on whether an enrichment method falls into the category of a univariate or multivariate approach, a variety of gene and/or gene set –level statistics can be selected. While multivariate approaches skip this step, in univariate approaches, a gene-level (i.e., local) statistic is first used to measure differences in gene expression levels and calculate scores for each gene in the dataset [8]. In Ackermann and Strimmer [35], the authors proposed a modular framework for enrichment analysis which they then used to compare 261 different variants on 10 simulated and 2 experimental datasets. In this study, the authors compared the following gene-level statistics: two-sample t-statistic, moderated t-statistic and Pearson correlation coefficient. Here, they concluded that the gene-level statistic used has little contribution to the overall rankings of the gene sets. By contrast, they noted that transformations of the gene-level statistic (i.e., squared values, ranks, local false discovery rate, binary transformations or none) do have an impact on the overall results, including the number of gene sets detected as significantly enriched. For instance, a transformation (e.g., absolute or squared value of the gene score) may be essential to detect gene sets that contain genes altered in opposite directions by a phenotype (i.e., both upregulated and downregulated).

Later investigations on the choice of gene-level statistic were conducted by Zyla *et al.* [68] where they evaluated the performance of 16 different ranking metrics in GSEA, a popular FCS univariate method, on 28 datasets. By an initial assignment of a target pathway to each dataset, the authors evaluated the overall sensitivity and false positive rate produced when different ranking metrics were employed. In contrast to [35], the authors concluded that the ranking metric does have a significant impact on results and highlighted four metrics (i.e., moderated Welch test, minimum significant difference, absolute value of signal-to-noise ratio, and Baumgartner-Weiss-Schindler test) that exhibited the best results with respect to overall sensitivity and false positive rate.

**Gene set – level statistics**

A number of gene set-level (i.e., global) statistics can be used for the calculation of the gene set score. This step is central to an enrichment analysis and ultimately determines whether a gene set is significantly enriched. In their comparative study, Ackermann and Strimmer [35] tested various gene-set level statistics, generally noting inconsistent performances across datasets (i.e., experimental and simulated). In their simulation, the authors found that the median of the transformed local statistic and the WRS test statistic can perform well when the competitive null hypothesis is tested and in the presence of outliers. Nonetheless, while the median was previously suggested in place of the mean, the authors note that the median, as well as the WRS test, may generate fewer overall results. A subsequent study by Hung *et al.* [69] also evaluated the impact of the global statistic on results by proposing a metric to assess the degree to which a gene set detected by a specific method can be reproduced by others. Here, the authors found that the WRS and Weighted Kolmogorov Smirnov (WKS) tests were able to obtain high scores on this metric and both are thus likely to be able to cover predictions by other evaluated global statistics.

**Multiple testing correction**

Lastly, another important methodological consideration includes the incorporation of corrections for multiple testing. A common scenario is one in which there are many gene sets for comparisons and few samples. As gene set collections can contain hundreds or even thousands of gene sets that are tested, type I errors can increase correspondingly to the number of tests [70]. While corrections are typically applied, depending on the dataset, enrichment method, and by extension, the model assumed by a given method, it may be advisable to ease or altogether forego multiple testing correction in certain cases, such as when a method is particularly conservative and is expected to yield little to no gene sets as significant [2].

### 3.5. Approaches that combine enrichment methods

Given the vast variety of enrichment methods, often with tunable settings, hundreds of methods and variants are at the disposal of life science researchers. As results can acutely vary according to the method selected, such a broad variability has prompted the development of tools to conduct enrichment analysis in concert. While the techniques to do so can differ, generally a consensus is taken across several methods to determine the final set of pathways that are interesting in some statistically significant way.

One example is the Ensemble of Genes Set Enrichment Analyses (EGSEA), an R package designed to combine the results of 12 distinct non topology-based methods by using different statistical methods to calculate a unified score for each gene set [71]. EGSEA has been implemented for RNA-seq data and leverages three gene set databases (i.e., MSigDB, KEGG, and GeneSetDB [72] ). Similarly, an equivalent R package, EnrichmentBrowser, runs several topology and non-topology based methods in parallel, before results are combined and gene sets are given new rankings, based on a defined ranking and combination function [73]. Väremo *et al.* [74] provide the Piano R package which incorporates variations of the global statistic and can be used to conduct enrichment analyses for microarray and RNA-seq data through a variety of methods to obtain consensus results.

More recent developments include the decoupleR package [75], made available for similar purposes, yet expanding upon existing ensemble approaches through its adaptation to bulk, single-cell as well as spatial *omics* data, and the CPA (Consensus Pathway Analysis) web application, enabling non-bioinformatician users to conduct enrichment analyses on multiple methods and databases [76]. In the latter case, consensus pathway analyses can be conducted on eight disparate enrichment methods, using KEGG, GO or user uploaded gene set collections.

Finally, Ai and Kong [77] demonstrate an ML-based approach, Combined Gene set analysis incorporating Prioritization and Sensitivity (CGPS), to combine the *p*-values and ranks of the results of nine enrichment methods to train a support vector machine (SVM) that outputs a consensus score. The authors found that CGPS yielded a better performance than the two previously mentioned methods (i.e., EGSEA and EnrichmentBrowser) in identifying relevant gene sets.

## 4. Impact of pathway database and gene set size

While variations of enrichment methods have been among the most studied factors that influence the results of an enrichment analysis, there are several other considerations to be made in the design of an experiment to ensure biologically meaningful results. In this section, we introduce studies, including notable benchmarks, that have investigated the impact of additional factors on the results of enrichment analysis, such as database choice and pathway size.

One of the most critical factors the results of an enrichment analysis can hinge upon is the choice of a reference pathway database(s). It is common practice for researchers to solely rely upon a single database for an enrichment analysis which can be due, in part, to a researcher's preferences, the popularity of a particular database or its ease of usage, among other factors. Indeed, we observed that the majority of studies which benchmarked the performance of enrichment methods **(Table 1)** were almost always conducted on a single database, and that too, primarily KEGG.

A first investigation on the importance of selecting a collection of gene sets was performed by Bateman *et al.* [78]. In this study, the authors demonstrated how the seven standard collections housed within MSigDB yielded different results when conducting GSEA within the context of a drug response cancer dataset. Among other findings, the results of this study indicated that some collections were able to yield a significantly larger number of enriched pathways relevant to the studied phenotype than others. Furthermore, the authors argued that the choice of gene set collections should not be made arbitrarily as certain gene sets may be more or less suitable for a particular dataset than others. In a recent study by Wieder *et al.* [79] on best practices for the popular ORA method on metabolomics data, the authors also found that the results of pathway analysis substantially differed based on the choice of pathway database (i.e., KEGG, Reactome and BioCyc [80]).

Similar conclusions were drawn in our previous work [81], in which we evaluated whether enrichment results are in consensus for any given pathway that can be found across three major pathway databases (i.e., KEGG, Reactome, and WikiPathways [82]) and multiple enrichment methods. Our study revealed the advantages of combining multiple databases by using equivalent pathway mappings, demonstrating that an integrative resource can yield more consistent results than an individual one. Overall, these studies demonstrate the importance of database choice, a crucial factor given the differences in coverage across databases [83-84]. Finally, we would also like to note the importance of database size as the total number of pathways present in a database has an influence when multiple correction methods are applied.

An additional factor that is related to database choice is gene set (pathway) size, corresponding to the number of genes within a gene set for enrichment methods that do not consider pathway topology, or the number of nodes (genes) and edges for those that do consider it. The effect of pathway size has recently been studied in Karp *et al.* [85] by comparing the significance of six equivalent pathway definitions from KEGG and EcoCyc [86]. Given the differences in the average size of a pathway across the two databases (i.e., KEGG pathways are significantly larger than their respective homologues in EcoCyc), the authors investigated the degree to which size could influence results, finding that pathway size can have a stronger effect than the statistical corrections used. Furthermore, the authors found that KEGG pathways required up to two times as many significant genes in order to attain the same *p*-value as their EcoCyc counterparts.

Notably, size differences between equivalent pathways have not only been examined for these two databases but also across other major resources, such as Reactome, and WikiPathways [84]. In this work, the authors argue that using pathway definitions which span across several biological processes (e.g., signal transduction) can lead to misinterpretations as when these pathways are enriched, it is difficult to construe whether this implicates all or only a subset of the pathway. These broadly defined pathways can also be less informative, contributing little in terms of novelty to the overall understanding of the distinctions between the phenotypes under study. Nonetheless, smaller pathways can lead to exceedingly long results and overly-strict multiple testing corrections [4].

Possible solutions for mitigating the impact of gene set size on results are defining the minimum and maximum number of genes within a gene set (e.g., between 10 and 500), careful consideration of the enrichment analysis method selected **(see Section 2.2)**, as well as addressing redundancies within gene sets, as proposed by Simillion *et al.* [87]. In their approach, the authors suggest discarding significant gene sets that overlap with others in order to ensure that the enrichment of a particular pathway is not a result of the overlay.

While database choice and pathway size are two critical factors to consider, we foresee several approaches to offset the challenges they create. In the case of database choice, a study by Maleki *et al.* [88] proposed two simple metrics (i.e., permeability and maximum achievable coverage scores) to assess the degree of overlap between a gene list of relevance and all gene sets within a database. The goal of these metrics is to provide an intuition of whether or not the genes of a phenotype under investigation are well covered by a particular database. Thus, the authors argue that this approach can reduce database bias and arbitrary database selection as the two scores can guide users to rationally decide upon the most appropriate database.

Another solution that we propose is that the enrichment results generated from a reference database could be validated against an additional database using equivalent pathway mappings across them. By leveraging pathway mappings, one can assess the similarity between the results obtained from different databases (i.e., reference and "validation" database) to confirm whether they are in consensus, or re-evaluate them if they are not. In earlier work, we leveraged this technique by generating equivalent pathway mappings across four pathway databases [89]. A web tool (i.e., DecoPath) subsequently enables users to evaluate similarities and differences at the gene and pathway level for a given pathway across databases and enrichment methods. For instance, a particular pathway in one database can have a slightly different gene set than the same pathway in another database, which can ultimately explain why a pathway is detected as significantly enriched in one database but not in another.

Similarly, pathway mappings can also be employed to systematically study the impact of pathway size on results. Here, one could leverage hierarchical mappings (i.e., pathway A *isPartOf* pathway B) from pathway ontologies to evaluate whether related pathways are similarly enriched. Although a pathway ontology was earlier proposed by [90], it has neither been adopted by or linked to any major database. Instead, each database utilizes its own pathway terminology, though some databases such as Reactome and GO also incorporate a hierarchical organization within their schema. In fact, Reactome recently adopted such an approach to facilitate the interpretation of enrichment analyses through implementing ReacFoam, a visualization for navigating through its pathway hierarchy and exploring the degree of enrichment of pathways at different levels.

The growth of biomedical literature is reflected in pathway databases as their pathway definitions change over time. A study by Wadi *et al*. [91] demonstrated the impact of outdated pathway definitions in several web-based tools as well as highlighted that the number of pathways/biological processes doubled in 7 years (2009-2016) in major resources such as Reactome and GO. Furthermore, it revealed that the majority of the studies analyzed were conducted using outdated pathway definitions, constituting a major issue as the results presented in such studies could have potentially changed. We believe this problem can be partially mitigated if users are alerted by pathway enrichment tools when the underlying pathway database(s) has not been recently updated. Furthemore, updating the information from pathway databases in a tool has been greatly simplified by the APIs and services offered by major resources such as Reactome, GO, and WikiPathways. Finally, we encourage researchers to include both the version of the database(s) used in the analysis as well as the version of the tool(s) employed.

# 5. Impact of additional factors on enrichment analysis and possible future benchmarks

While the factors mentioned thus far have each been benchmarked with regard to their impact on pathway enrichment results, there exist other factors that have not yet been explored in detail. Firstly, at a more granular level, individual genes can also have an impact on results. A study by Ballouz *et al.* [92] raised the challenges associated with annotation bias and redundancies in gene sets. The annotation of a single gene to many functions (i.e., multifunctional genes) can potentially confound the results of a pathway analysis as these genes may result in a sizeable number of enriched pathways that are largely irrelevant. For example, several pathways with multifunctional genes may be considered enriched in the results, though the enrichment of these pathways could be due to the presence of multifunctional genes rather than the relevance of the pathway to the phenotype of interest. One approach the authors propose to control for this effect is by performing repeated runs of the analysis while removing the topmost multifunctional genes in the dataset, in order to identify the most robust pathways.

Furthermore, other ways to reduce the effect of multifunctional genes can include assigning weights to genes based on their promiscuity, though this approach might also have drawbacks.

A second factor that has not yet been investigated which is related both to database updates and choice is the size of a database measured by the number of pathways. This factor is not only important due to its correlation with the coverage of biological processes, but also because the size of the database can influence the significance of the results when correcting for multiple testing. As a consequence, depending on the size of a database, the same pathway in one database may or may not be enriched in another after applying multiple testing correction. This is often the case when comparing popular databases, such as KEGG and Reactome, whose number of pathways can differ by an order of magnitude.

Finally, we would like to note that there are other interesting factors which could potentially be analyzed in the future. Firstly, for topology-based methods, the particular network structure of some pathways may make them more susceptible to enrichment than others given the topological differences identified by [93]. Thus, one future possible benchmark could investigate the effect of network sparsity on pathway enrichment, or if the presence of hubs correlates with greater enrichment. Secondly, another factor to evaluate is the degree to which a bias towards certain indications in pathway knowledge influences results. For example, there is an over-representation of interactions characterized in widely studied indication areas, such as cancer [94-95], and thus, pathways containing these interactions may appear in the results of enrichment, while possessing little relevance to the studied phenotype. To investigate this factor, resources such as BioGrid [96] where protein-protein interactions are annotated with experimental metadata can be leveraged, since the majority of databases do not provide information on the provenance supporting each interaction.

# 6. Discussion

The last decade has seen an explosion in the usage of pathway enrichment analysis, spearheaded by both an abundance in the volume of available data and the interpretive power of these analyses [10]. Prompted by a wide range of available enrichment methods and pathway resources, several comparative studies have evaluated how different factors can influence the results of such an analysis. Here, we have reviewed the findings of these studies in order to provide a comprehensive overview on the impact of these factors. Furthermore, we have suggested possible approaches to overcome some of the limitations discussed as well as possibilities for additional benchmark studies on other, under studied factors.

In the first section of this review, we have outlined and summarized the results of 12 comparative studies that have investigated differences across pathway enrichment methods. Many of these studies have specifically focused on the performance of individual methods on popular metrics (e.g., prioritization, sensitivity, and specificity), keeping in mind that without gold standards to conclude whether the results from any given method are biologically sound, objective evaluations can be difficult to make. Overall, we have found many inconsistencies in the performance of methods across metrics as well as across studies. While there is no consensus across studies on whether a specific method outperforms others, we have reported trends we have observed regarding the top performing methods for each metric.

Though we note that the performances of the majority of methods on these and other metrics is inconclusive, whether a particular method is a reasonable choice for a certain use case can depend on a number of factors, such as the goal of the experiment, the dataset in question or particulars of the gene set collection. Nevertheless, tradeoffs between performance on certain metrics can be important considerations in the selection of a method. For example, given a dataset where changes in differential gene expression between experimental groups are subtle, a highly sensitive method can increase the likelihood of detecting a signal. Thus, a large number of gene sets which are truly significant can be identified, essentially ruling out nearly all gene sets that are not detected, albeit at the expense of producing a greater number of false positives. If, however, changes in differential gene expression between experimental groups are generally more pronounced, a method ranked high in specificity may be preferable to preclude the detection of too many gene sets which can complicate interpretation.

We have also examined comparative studies that have evaluated the differences between distinct categories of enrichment methods, such as how the null hypothesis is formulated (e.g., self-contained and competitive) and the sampling unit is defined (i.e., gene or subject), noting that the selection of one category of methods over another can have serious repercussions on the fraction of gene sets that are significant and their ranks. Additionally, a major categorical distinction is drawn between topology and non-topology -based methods which have been reviewed in several benchmarks. We have found that, though topology-based approaches are more advanced, for some methods, the removal of topological information yields no differences in results, for other methods it can improve results, and several are constrained in that they only cater to KEGG pathways (or pathways in an equivalent format). Finally, we reviewed studies that have assessed the influence of particular, modular aspects of a typical enrichment analysis as well as outlined additional aspects one must be cognizant of that can affect the behaviour of a given method, which ultimately reflects in the overall results of an analysis.

We have reviewed several other factors apart from enrichment methods, such as pathway size and database choice. Notably, the latter can be subjective, with both researcher preferences and distinct research goals taking precedence over set guidelines [4]. However, we have outlined approaches that leverage pathway mappings to mitigate the effect of these factors. An additional aspect discussed in this review are the lack of regular updates to enrichment tools which reflect updates made to pathway databases. Fortunately, this issue has, at least, partially been addressed by the adoption of API services by major pathway resources. Nevertheless, the amount of literature published on a daily basis continues to grow, making the task of maintaining up-to-date pathway definitions difficult, particularly for public and academic resources. Thus, we envisage that the path forward to address this shortfall is to improve interoperability across databases via mappings [84] or through the use of common database formats [97].

Finally, we would like to mention possible future benchmarks beyond the ones we have previously proposed. Firstly, we believe that future benchmarks would benefit from the existence of a gold standard prioritization approach, for instance, one that leverages well-established pathway-disease associations from genetic disorders, similar to the assessment proposed by Nguyen *et al.* [9], which exploits knockout datasets. Secondly, given the rise of multi-*omics* datasets, we anticipate the development of enrichment methods that operate on other modalities beyond mRNA data, such as metabolomics. Finally, we foresee that the insights gained from multi-*omics* experiments will also be reflected in pathway definitions in two ways: i) the appearance of "dynamic pathways" (i.e., contextualized pathways representing particular pathway states as opposed to the currently available general, static diagrams), and ii) a shift from traditional gene sets to sets of multimodal biological entities.

# 7. Conclusion

In conclusion, the effect of various factors on pathway enrichment analysis are apparent. Numerous studies have demonstrated how variations in the design of an enrichment analysis can lead to altogether different findings. At the extremes, comparative studies have shown how certain experimental setups can detect either all or no gene sets as interesting in some statistically significant way. We summarize the key findings of studies reviewed herein as follows:

**Data preprocessing:** Careful selection of the approach to preprocess data, including normalization, is crucial in ensuring fair comparisons across samples are made and to minimize the propagation of errors in the downstream analysis.

**Formulation of null hypothesis and significance assessment:** One must be cognizant of how the null hypothesis is formulated (i.e., competitive or self-contained) as methods categorized into one or another approach behave differently in terms of the fraction of gene sets reported as significant, as well as their sensitivity to gene set size, sample size, and sample heterogeneity. Self-contained methods also tend to have greater power than competitive methods and careful consideration should be made taking into account the proportion of genes that are differentially expressed in the dataset. Similarly, in order to calculate a $p$-value for each gene set, one must bear in mind that disparate approaches can impact the results of an enrichment analysis, and depending on the approach taken, introduce false positives.

**Pathway and sample size considerations:** Certain enrichment methods have been observed to be more or less robust to pathway and sample size than certain others. Sensitive methods may detect larger gene sets as significantly enriched and their sensitivity can be tied with whether they are competitive or self-contained methods. Not surprisingly, a methods' performance tends to deteriorate with decreasing sample size, although some methods are more robust on this factor than others.

**Gene and gene set – level statistics and transformation:** Although differing studies have reached different conclusions on the importance of this factor, the careful choice of a gene-level statistic is at least essential when also paired with an effective transformation strategy. At the heart of an enrichment analysis is the gene set – level statistic to determine which gene sets are significantly enriched. Benchmark studies have proposed methods that may be suitable for its selection.

**Topology vs non-topology -based methods:** Topology-based methods are intuitively more advanced than non-topology -based ones. Incorporation of topological information tends to improve the ranks and $p$-values of relevant pathways for some topology-based methods, yet this may not be the case for all. Nonetheless, some topology-based methods are limited or at least partial to specific pathway databases.

**Choice of gene set collection or pathway database:** The selection of one gene set collection over another can lead to different results. Some collections or databases may be more suitable than others for a given dataset. The selection of a database is complicated by variable definitions of pathway boundaries as well as by redundancies and outdated pathway definitions.

The errors from these steps that propagate through an enrichment analysis may be inconsequential at best and misleading at worst. Although there is no singular method or gene set collection/pathway database which is advisable for enrichment analysis over all others, well informed choices can be made and solutions to mitigate the impact of various factors are available. Furthermore, recently, many ensemble approaches have been developed so that users can benefit from multiple databases and/or methods.

**Funding**

This work was developed in the Fraunhofer Cluster of Excellence "Cognitive Internet Technologies".

**Authors' Contributions**

SM and DDF wrote the manuscript. ATK and MHA reviewed the manuscript.

All authors have read and approved the final manuscript.

**Competing interests**

DDF received salary from Enveda Biosciences.

# References

1. Khatri, P., Sirota, M., and Butte, A. J. (2012). Ten years of pathway analysis: current approaches and outstanding challenges. *PLoS Comput Biol, 8*(2), e1002375. https://doi.org/10.1371/journal.pcbi.1002375

2. Geistlinger, L., Csaba, G., Santarelli, M., Ramos, M., Schiffer, L., Turaga, N., *et al.* (2020). Toward a gold standard for benchmarking gene set enrichment analysis. *Briefings in bioinformatics, 22*(1), 545-556. https://doi.org/10.1093/bib/bbz158

3. Ihnatova, I., Popovici, V., and Budinska, E. (2018). A critical comparison of topology-based pathway analysis methods. *PloS one*, *13*(1), e0191154. https://doi.org/10.1371/journal.pone.0191154

4. Reimand, J., Isserlin, R., Voisin, V., Kucera, M., Tannus-Lopes, C., Rostamianfar, A., *et al*. (2019). Pathway enrichment analysis and visualization of omics data using g:Profiler, GSEA, Cytoscape and EnrichmentMap. *Nature protocols, 14*(2), 482-517. https://doi.org/10.1038/s41596-018-0103-9

5. Kanehisa, M., Furumichi, M., Sato, Y., Ishiguro-Watanabe, M., and Tanabe, M. (2021). KEGG: integrating viruses and cellular organisms. *Nucleic Acids Research, 49*(D1), D545-D551. https://doi.org/10.1093/nar/gkaa970

6. Fabregat, A., Jupe, S., Matthews, L., Sidiropoulos, K., Gillespie, M., Garapati, P., *et al.* (2018). The Reactome pathway Knowledgebase. *Nucleic acids research*, *46*(D1):D649-D655. https://doi.org/10.1093/nar/gkx1132

7. The Gene Ontology Consortium. (2021). The Gene Ontology resource: enriching a GOld mine. *Nucleic Acids Research, 49*(D1), D325-D334. https://doi.org/10.1093/nar/gkaa1113

8. Maleki, F., Ovens, K., Hogan, D. J., and Kusalik, A. J. (2020). Gene set analysis: challenges, opportunities, and future research. *Frontiers in genetics, 11*(654). https://doi.org/10.3389/fgene.2020.00654

9. Nguyen, T. M., Shafi, A., Nguyen, T., and Draghici, S. (2019). Identifying significantly impacted pathways: a comprehensive review and assessment. *Genome biology, 20*(1), 1-15. https://doi.org/10.1186/s13059-019-1790-4

10. Xie, C., Jauhari, S., and Mora, A. (2021). Popularity and performance of bioinformatics software: the case of gene set analysis. *BMC bioinformatics, 22*(1), 1-16. https://doi.org/10.1186/s12859-021-04124-5

11. Massey Jr, F. J. (1951). The Kolmogorov-Smirnov test for goodness of fit. *Journal of the American Statistical Association, 46*(253), 68-78. https://doi.org/10.1080/01621459.1951.10500769

12. Wilcoxon, F. (1992). Individual comparisons by ranking methods. In Breakthroughs in statistics (pp. 196-202). Springer, New York, NY. https://doi.org/10.2307/3001968

13. Bayerlová, M., Jung, K., Kramer, F., Klemm, F., Bleckmann, A., and Beißbarth, T. (2015). Comparative study on gene set and pathway topology-based enrichment methods. *BMC bioinformatics*, *16*(1), 334. https://doi.org/10.1186/s12859-015-0751-5

14. Jaakkola, M. K., and Elo, L. L. (2016). Empirical comparison of structure-based pathway methods. B*riefings in bioinformatics*, 17(2), 336-345. https://doi.org/10.1093/bib/bbv049

15. Ma, J., Shojaie, A., and Michailidis, G. (2019). A comparative study of topology-based pathway enrichment analysis methods. *BMC bioinformatics*, *20*(1), 1-14. https://doi.org/10.1186/s12859-019-3146-1

16. Maleki, F., Ovens, K. L., Hogan, D. J., Rezaei, E., Rosenberg, A. M., and Kusalik, A. J. (2019a). Measuring consistency among gene set analysis methods: A systematic study. *Journal of bioinformatics and computational biology, 17*(05), 1940010. https://doi.org/10.1142/S0219720019400109

17. Liberzon, A., Subramanian, A., Pinchback, R., Thorvaldsdóttir, H., Tamayo, P., and Mesirov, J. P. (2011). Molecular signatures database (MSigDB) 3.0. *Bioinformatics, 27*(12), 1739-1740. https://doi.org/10.1093/bioinformatics/btr260

18. Mathur, R., Rotroff, D., Ma, J., Shojaie, A., and Motsinger-Reif, A. (2018). Gene set analysis methods: a systematic comparison. *BioData mining, 11*(1), 1-19. https://doi.org/10.1186/s13040-018-0166-8

19. Dutta, B., Wallqvist, A., and Reifman, J. (2012). PathNet: a tool for pathway analysis using topological information. *Source code for biology and medicine, 7*(1), 1-12. https://doi.org/10.1186/1751-0473-7-10

20. Gu, Z., Liu, J., Cao, K., Zhang, J., and Wang, J. (2012). Centrality-based pathway enrichment: a systematic approach for finding significant pathways dominated by key genes. *BMC systems biology, 6*(1), 1-13. https://doi.org/10.1186/1752-0509-6-56

21. Gu, Z., and Wang, J. (2013). CePa: an R package for finding significant pathways weighted by multiple network centralities. *Bioinformatics, 29*(5), 658-660. https://doi.org/10.1093/bioinformatics/btt008

22. Rahmatallah, Y., Emmert-Streib, F., and Glazko, G. (2016). Gene set analysis approaches for RNA-seq data: performance evaluation and application guideline. *Briefings in bioinformatics, 17*(3), 393-407. https://doi.org/10.1093/bib/bbv069

23. Tarca, A. L., Bhatti, G., and Romero, R. (2013). A comparison of gene set analysis methods in terms of sensitivity, prioritization and specificity. *PloS one, 8*(11), e79217. https://doi.org/10.1371/journal.pone.0079217

24. Zyla, J., Marczyk, M., and Polanska, J. (2017a). Reproducibility of finding enriched gene sets in biological data analysis. *In International Conference on Practical Applications of Computational Biology & Bioinformatics* (pp. 146-154). https://doi.org/10.1007/978-3-319-60816-7_18

25. Zyla, J., Marczyk, M., Domaszewska, T., Kaufmann, S. H., Polanska, J., and Weiner 3rd, J. (2019). Gene set enrichment for reproducible science: comparison of CERNO and eight other algorithms. *Bioinformatics*, *35*(24), 5146-5154. https://doi.org/10.1093/bioinformatics/btz447

26. Michaud, J., Simpson, K. M., Escher, R., Buchet-Poyau, K., Beissbarth, T., Carmichael, C., *et al.* (2008). Integrative analysis of RUNX1 downstream pathways and target genes. *BMC genomics, 9*(1), 1-17. https://doi.org/10.1186/1471-2164-9-363

27. Goeman, J. J., Van De Geer, S. A., De Kort, F., and Van Houwelingen, H. C. (2004). A global test for groups of genes: testing association with a clinical outcome. *Bioinformatics, 20*(1), 93-99. https://doi.org/10.1093/bioinformatics/btg382

28. Tomfohr, J., Lu, J., and Kepler, T. B. (2005). Pathway level analysis of gene expression using singular value decomposition. *BMC bioinformatics, 6*(1), 1-11. https://doi.org/10.1186/1471-2105-6-225

29. Tarca, A. L., Draghici, S., Khatri, P., Hassan, S. S., Mittal, P., Kim, J. S., *et al.* (2009). A novel signaling pathway impact analysis. *Bioinformatics, 25*(1), 75-82. https://doi.org/10.1093/bioinformatics/btn577

30. Wu, D., and Smyth, G. K. (2012). Camera: a competitive gene set test accounting for inter-gene correlation. *Nucleic acids research, 40*(17), e133-e133. https://doi.org/10.1093/nar/gks461

31. Efron, B., and Tibshirani, R. (2007). On testing the significance of sets of genes. *The annals of applied statistics, 1*(1), 107-129. https://doi.org/10.1214/07-AOAS101

32. Tarca, A. L., Draghici, S., Bhatti, G., and Romero, R. (2012). Down-weighting overlapping genes improves gene set analysis. *BMC bioinformatics, 13*(1), 1-14. https://doi.org/10.1186/1471-2105-13-136

33. Subramanian, A., Tamayo, P., Mootha, V. K., Mukherjee, S., Ebert, B. L., Gillette, M. A., *et al.* (2005). Gene set enrichment analysis: a knowledge-based approach for interpreting genome-wide expression profiles. *Proceedings of the National Academy of Sciences, 102*(43), 15545-15550. https://doi.org/10.1073/pnas.0506580102

34. Goeman, J. J., and Bühlmann, P. (2007). Analyzing gene expression data in terms of gene sets: methodological issues. *Bioinformatics, 23*(8), 980-987. https://doi.org/10.1093/bioinformatics/btm051

35. Ackermann, M., and Strimmer, K. (2009). A general modular framework for gene set enrichment analysis. *BMC bioinformatics, 10*(1), 1-20. https://doi.org/10.1186/1471-2105-10-47

36. Tripathi, S., Glazko, G. V., and Emmert-Streib, F. (2013). Ensuring the statistical soundness of competitive gene set approaches: gene filtering and genome-scale coverage are essential. *Nucleic acids research, 41*(7), e82-e82. https://doi.org/10.1093/nar/gkt054

37. Wu, M. C., and Lin, X. (2009). Prior biological knowledge-based approaches for the analysis of genome-wide expression profiles using gene sets and pathways. *Statistical methods in medical research, 18*(6), 577-593. https://doi.org/10.1177/0962280209351925

38. Wu, D., Lim, E., Vaillant, F., Asselin-Labat, M. L., Visvader, J. E., and Smyth, G. K. (2010). ROAST: rotation gene set tests for complex microarray experiments. *Bioinformatics, 26*(17), 2176-2182. https://doi.org/10.1093/bioinformatics/btq401

39. Dinu, I., Potter, J. D., Mueller, T., Liu, Q., Adewale, A. J., Jhangri, G. S., *et al.* (2007). Improving gene set analysis of microarray data by SAM-GS. *BMC bioinformatics, 8*(1), 1-13. https://doi.org/10.1186/1471-2105-8-242

40. Maciejewski, H. (2014). Gene set analysis methods: statistical models and methodological differences. *Briefings in bioinformatics, 15*(4), 504-518. https://doi.org/10.1093/bib/bbt002

41. Barry, W. T., Nobel, A. B., and Wright, F. A. (2005). Significance analysis of functional categories in gene expression studies: a structured permutation approach. *Bioinformatics, 21*(9), 1943-1949. https://doi.org/10.1093/bioinformatics/bti260

42. Nam, D. (2017). Effect of the absolute statistic on gene-sampling gene-set analysis methods. *Statistical methods in medical research, 26*(3), 1248-1260. https://doi.org/10.1177/0962280215574014

43. Barry, W. T., Nobel, A. B., and Wright, F. A. (2008). A statistical framework for testing functional categories in microarray data. *The Annals of Applied Statistics, 2*(1), 286-315. https://doi.org/10.1214/07-AOAS146

44. Tamayo, P., Steinhardt, G., Liberzon, A., and Mesirov, J. P. (2016). The limitations of simple gene set enrichment analysis assuming gene independence. *Statistical methods in medical research, 25*(1), 472-487. https://doi.org/10.1177/0962280212460441

45. Irizarry, R. A., Wang, C., Zhou, Y., and Speed, T. P. (2009). Gene set enrichment analysis made simple. *Statistical methods in medical research, 18*(6), 565-575. https://doi.org/10.1177/0962280209351908

46. Kim, S. Y., and Volsky, D. J. (2005). PAGE: parametric analysis of gene set enrichment. *BMC bioinformatics, 6*(1), 1-12. https://doi.org/10.1186/1471-2105-6-144

47. Tian, L., Greenberg, S. A., Kong, S. W., Altschuler, J., Kohane, I. S., and Park, P. J. (2005). Discovering statistically significant pathways in expression profiling studies. *Proceedings of the National Academy of Sciences*, 102(38), 13544-13549. https://doi.org/10.1073/pnas.0506577102

48. Maleki, F., Ovens, K., McQuillan, I., and Kusalik, A. J. (2018). Sample size and reproducibility of gene set analysis. *IEEE International Conference on Bioinformatics and Biomedicine,* 122-129. https://doi.org/10.1109/BIBM.2018.8621462

49. Ritchie, M. E., Phipson, B., Wu, D. I., Hu, Y., Law, C. W., Shi, W., and Smyth, G. K. (2015). limma powers differential expression analyses for RNA-sequencing and microarray studies. *Nucleic Acids Research, 43*(7), e47-e47. https://doi.org/10.1093/nar/gkv007

50. Zahn, J. M., Sonu, R., Vogel, H., Crane, E., Mazan-Mamczarz, K., Rabkin, R., *et al.* (2006). Transcriptional profiling of aging in human muscle reveals a common aging signature. *PLoS genetics, 2*(7), e115. https://doi.org/10.1371/journal.pgen.0020115

51. Massa, M. S., Chiogna, M., and Romualdi, C. (2010). Gene set analysis exploiting the topology of a pathway. *BMC systems biology, 4*(1), 1-15. https://doi.org/10.1186/1752-0509-4-121

52. Martini, P., Sales, G., Massa, M. S., Chiogna, M., and Romualdi, C. (2013). Along signal paths: an empirical gene set approach exploiting pathway topology. *Nucleic acids research, 41*(1), e19-e19. https://doi.org/10.1093/nar/gks866

53. Ibrahim, M. A. H., Jassim, S., Cawthorne, M. A., and Langlands, K. (2012). A topology-based score for pathway enrichment. *Journal of Computational Biology, 19*(5), 563-573. https://doi.org/10.1089/cmb.2011.0182

54. Hänzelmann, S., Castelo, R., and Guinney, J. (2013). GSVA: gene set variation analysis for microarray and RNA-seq data. *BMC bioinformatics, 14*(1), 1-15. https://doi.org/10.1186/1471-2105-14-7

55. Luo, W., Friedman, M. S., Shedden, K., Hankenson, K. D., & Woolf, P. J. (2009). GAGE: generally applicable gene set enrichment for pathway analysis. *BMC bioinformatics, 10*(1), 1-17. https://doi.org/10.1186/1471-2105-10-161

56. Grant, G. R., Manduchi, E., and Stoeckert Jr, C. J. (2007). Analysis and management of microarray gene expression data. *Current protocols in molecular biology, 77*(1), 19-6. https://doi.org/10.1002/0471142727.mb1906s77

57. Irizarry, R. A., Hobbs, B., Collin, F., Beazer-Barclay, Y. D., Antonellis, K. J., Scherf, U., and Speed, T. P. (2003). Exploration, normalization, and summaries of high density oligonucleotide array probe level data. *Biostatistics, 4*(2), 249-264. https://doi.org/10.1093/biostatistics/4.2.249

58. Wu, Z., Irizarry, R. A., Gentleman, R., Martinez-Murillo, F., and Spencer, F. (2004). A model-based background adjustment for oligonucleotide expression arrays. *Journal of the American Statistical Association, 99*(468), 909-917. https://doi.org/10.1198/016214504000000683

59. Huber, W., Von Heydebreck, A., Sültmann, H., Poustka, A., and Vingron, M. (2002). Variance stabilization applied to microarray data calibration and to the quantification of differential expression. *Bioinformatics, 18*(suppl_1), S96-S104. https://doi.org/10.1093/bioinformatics/18.suppl_1.S96

60. Irizarry, R. A., Wu, Z., and Jaffee, H. A. (2006). Comparison of Affymetrix GeneChip expression measures. *Bioinformatics, 22*(7), 789-794. https://doi.org/10.1093/bioinformatics/btk046

61. Dillies, M. A., Rau, A., Aubert, J., Hennequet-Antier, C., Jeanmougin, M., Servant, N., *et al.* (2013). A comprehensive evaluation of normalization methods for Illumina high-throughput RNA sequencing data analysis. *Briefings in bioinformatics, 14*(6), 671-683. https://doi.org/10.1093/bib/bbs046

62. Anders, S., and Huber, W. (2010) Differential expression analysis for sequence count data. *Genome Biology 11*(10), R106. https://doi.org/10.1186/gb-2010-11-10-r106

63. Robinson, M. D., and Oshlack, A. (2010). A scaling normalization method for differential expression analysis of RNA-seq data. *Genome biology, 11*(3), 1-9. https://doi.org/10.1186/gb-2010-11-3-r25

64. Mortazavi, A., Williams, B. A., McCue, K., Schaeffer, L., and Wold, B. (2008). Mapping and quantifying mammalian transcriptomes by RNA-Seq. *Nature methods, 5(*7), 621-628. https://doi.org/10.1038/nmeth.1226

65. Holland, C. H., Tanevski, J., Perales-Patón, J., Gleixner, J., Kumar, M. P., Mereu, E., *et al.* (2020). Robustness and applicability of transcription factor and pathway analysis tools on single-cell RNA-seq data. *Genome biology, 21*(1), 1-19. https://doi.org/10.1186/s13059-020-1949-z

66. Mora, A. (2020). Gene set analysis methods for the functional interpretation of non-mRNA data—Genomic range and ncRNA data. *Briefings in bioinformatics, 21*(5), 1495-1508. https://doi.org/10.1093/bib/bbz090

67. Maksimovic, J., Oshlack, A., and Phipson, B. (2021). Gene set enrichment analysis for genome-wide DNA methylation data. *Genome biology, 22*(1), 1-26. https://doi.org/10.1186/s13059-021-02388-x

68. Zyla, J., Marczyk, M., Weiner, J., and Polanska, J. (2017b). Ranking metrics in gene set enrichment analysis: do they matter?. *BMC bioinformatics, 18*(1), 1-12. https://doi.org/10.1186/s12859-017-1674-0

69. Hung, J. H., Yang, T. H., Hu, Z., Weng, Z., and DeLisi, C. (2012). Gene set enrichment analysis: performance evaluation and usage guidelines. *Briefings in bioinformatics, 13*(3), 281-291. https://doi.org/10.1093/bib/bbr049

70. Korthauer, K., Kimes, P. K., Duvallet, C., Reyes, A., Subramanian, A., Teng, M., *et al.* (2019). A practical guide to methods controlling false discoveries in computational biology. *Genome biology, 20*(1), 1-21. https://doi.org/10.1186/s13059-019-1716-1

71. Alhamdoosh, M., Ng, M., Wilson, N. J., Sheridan, J. M., Huynh, H., Wilson, M. J., and Ritchie, M. E. (2017). Combining multiple tools outperforms individual methods in gene set enrichment analyses. *Bioinformatics*, *33*(3), 414-424. https://doi.org/10.1093/bioinformatics/btw623

72. Araki, H., Knapp, C., Tsai, P., and Print, C. (2012). GeneSetDB: a comprehensive meta-database, statistical and visualisation framework for gene set analysis. *FEBS open bio*, *2*, 76-82. https://doi.org/10.1016/j.fob.2012.04.003

73. Geistlinger, L., Csaba, G., and Zimmer, R. (2016). Bioconductor's EnrichmentBrowser: seamless navigation through combined results of set-& network-based enrichment analysis. *BMC bioinformatics, 17*(1), 1-11. https://doi.org/10.1186/s12859-016-0884-1

74. Väremo, L., Nielsen, J., and Nookaew, I. (2013). Enriching the gene set analysis of genome-wide data by incorporating directionality of gene expression and combining statistical hypotheses and methods. *Nucleic acids research, 41*(8), 4378-4391. https://doi.org/10.1093/nar/gkt111

75. Badia-i-Mompel, P., Vélez, J., Braunger, J., Geiss, C., Dimitrov, D., Müller-Dott, S., *et al.* (2021). decoupleR: Ensemble of computational methods to infer biological activities from omics data. bioRxiv. https://doi.org/10.1101/2021.11.04.467271

76. Nguyen, H., Tran, D., Galazka, J. M., Costes, S. V., Beheshti, A., Petereit, J., *et al.* (2021). CPA: a web-based platform for consensus pathway analysis and interactive visualization. *Nucleic Acids Research*, gkab421, https://doi.org/10.1093/nar/gkab421

77. Ai, C., and Kong, L. (2018). CGPS: A machine learning-based approach integrating multiple gene set analysis tools for better prioritization of biologically relevant pathways. *Journal of genetics and genomics, 45*(9), 489-504. https://doi.org/10.1016/j.jgg.2018.08.002

78. Bateman, A. R., El-Hachem, N., Beck, A. H., Aerts, H. J., and Haibe-Kains, B. (2014). Importance of collection in gene set enrichment analysis of drug response in cancer cell lines. *Scientific reports, 4*, 4092. https://doi.org/10.1038/srep04092

79. Wieder, C., Frainay, C., Poupin, N., Rodríguez-Mier, P., Vinson, F., Cooke, J., *et al.* (2021). Pathway analysis in metabolomics: Recommendations for the use of over-representation analysis. *PLoS Comput Biol 17*(9): e1009105. https://doi.org/10.1371/journal.pcbi.1009105

80. Karp, P. D., Billington, R., Caspi, R., Fulcher, C. A., Latendresse, M., Kothari, A., *et al.* (2019). The BioCyc collection of microbial genomes and metabolic pathways. *Briefings in bioinformatics, 20*(4), 1085-1093. https://doi.org/10.1093/bib/bbx085

81. Mubeen, S., Hoyt, C. T., Gemünd, A., Hofmann-Apitius, M., Fröhlich, H., and Domingo-Fernández, D. (2019). The impact of pathway database choice on statistical enrichment analysis and predictive modeling. *Frontiers in genetics, 10,* 1203. https://doi.org/10.3389/fgene.2019.01203

82. Martens, M., Ammar, A., Riutta, A., Waagmeester, A., Slenter, D. N., Hanspers, K., *et al.* (2021). WikiPathways: connecting communities. *Nucleic Acids Research, 49*(D1), D613-D621. https://doi.org/10.1093/nar/gkaa1024

83. Stobbe, M. D., Houten, S. M., Jansen, G. A., van Kampen, A. H., and Moerland, P. D. (2011). Critical assessment of human metabolic pathway databases: a stepping stone for future integration. *BMC systems biology, 5*(1), 165. https://doi.org/10.1186/1752-0509-5-165

84. Domingo-Fernández, D., Hoyt, C. T., Bobis-Álvarez, C., Marin-Llao, J., and Hofmann-Apitius, M. (2018). ComPath: An ecosystem for exploring, analyzing, and curating mappings across pathway databases. *npj Systems Biology and Applications*, *4*(1):43. https://doi.org/10.1038/s41540-018-0078-8

85. Karp, P. D., Midford, P. E., Caspi, R., and Khodursky, A. (2021). Pathway size matters: the influence of pathway granularity on over-representation (enrichment analysis) statistics. *BMC genomics*, *22*(1), 1-11. https://doi.org/10.1186/s12864-021-07502-8

86. Keseler, I. M., Mackie, A., Santos-Zavaleta, A., Billington, R., Bonavides-Martínez, C., Caspi, R., *et al.* (2017). The EcoCyc database: reflecting new knowledge about Escherichia coli K-12. *Nucleic acids research*, *45*(D1), D543-D550. https://doi.org/10.1093/nar/gkw1003

87. Simillion, C., Liechti, R., Lischer, H. E., Ioannidis, V., and Bruggmann, R. (2017). Avoiding the pitfalls of gene set enrichment analysis with SetRank. *BMC bioinformatics*, *18*(1), 1-14. https://doi.org/10.1186/s12859-017-1571-6

88. Maleki, F., Ovens, K., McQuillan, I., Rezaei, E., Rosenberg, A. M., and Kusalik, A. J. (2019b). Gene set databases: a fountain of knowledge or a siren call? *Proceedings of the 10th ACM International Conference on Bioinformatics, Computational Biology and Health Informatics* (pp. 269-278). https://doi.org/10.1145/3307339.3342146

89. Mubeen, S., Bharadhwaj, V. S., Gadiya, Y., Hofmann-Apitius, M., and Domingo-Fernández, D. (2021). DecoPath: a web application for decoding pathway enrichment analysis. *NAR Genomics and Bioinformatics*, *3*(3), lqab087. https://doi.org/10.1093/nargab/lqab087

90. Petri, V., Jayaraman, P., Tutaj, M., Hayman, G. T., Smith, J. R., De Pons, J., *et al.* (2014). The pathway ontology–updates and applications. *Journal of biomedical semantics*, *5*(1), 1-12. https://doi.org/10.1186/2041-1480-5-7

91. Wadi, L., Meyer, M., Weiser, J., Stein, L. D., and Reimand, J., *et al.* (2016). Impact of outdated gene annotations on pathway enrichment analysis. *Nature methods*, *13*(9):705. https://doi.org/10.1038/nmeth.3963

92. Ballouz, S., Pavlidis, P., and Gillis, J. (2017). Using predictive specificity to determine when gene set analysis is biologically meaningful. *Nucleic acids research, 45*(4), e20-e20. https://doi.org/10.1093/nar/gkw957

93. Rubel, T., Singh, P., and Ritz, A. (2021). Stop Bickering! Reconciling Signaling Pathway Databases with Network Topologies. bioRxiv.

94. Reyes-Aldasoro, C. C. (2017). The proportion of cancer-related entries in PubMed has increased considerably; is cancer truly "The Emperor of All Maladies"?. *PloS one*, *12*(3), e0173671. https://doi.org/10.1371/journal.pone.0173671

95. Hanspers, K., Riutta, A., Summer-Kutmon, M., and Pico, A. R. (2020). Pathway information extracted from 25 years of pathway figures. *Genome biology*, *21*(1), 1-18. https://doi.org/10.1186/s13059-020-02181-2

96. Oughtred, R., Stark, C., Breitkreutz, B.J., Rust, J., Boucher, L., Chang, C., *et al.* (2019). The BioGRID interaction database: 2019 update. *Nucleic Acids Research*, *47*(D1), D529-D541. https://doi.org/10.1093/nar/gkw1102

97. Good, B. M., Van Auken, K., Hill, D. P., Mi, H., Carbon, S., Balhoff, J. P., *et al.* (2021). Reactome and the Gene Ontology: Digital convergence of data resources. *Bioinformatics*, btab325. https://doi.org/10.1093/bioinformatics/btab325

# Supplementary Tables

| Table | Description |
| --- | --- |
| STable 1 - Studies | Comparative studies included in review |
| STable 2 - Methods | Methods evaluated in comparative studies |
| STable 3 - Sensitivity | Method performance on sensitivity metric (True Positive Rate (TPR)/target pathway $p$-values) |
| STable 4 - Specificity | Method performance on specificity metric (True Negative Rate (TNR)) |
| STable 5 - Prioritization | Method performance on prioritization metric (target pathway rank) |
| STable 6 - Sample size robustness | Method performance on robustness to sample size |
| STable 7 - Gene set size robustness | Method performance on sensitivity to gene set size |
| STable 8 - Phenotype relevance | Method performance on ranking phenotype relevant gene sets |
| STable 9 - Accuracy | Method performance on accuracy |
| STable 10 - Additional measures | Additional measures comparing method performance |

# STable 1 - Studies

| Study # | Study | DOI |
|---|---|---|
| 1 | Bayerlová et al., 2015 | https://doi.org/10.1186/s12859-015-0751-5 |
| 2 | Geistlinger et al., 2021 | https://doi.org/10.1093/bib/bbz158 |
| 3 | Ihnatova et al., 2018 | https://doi.org/10.1371/journal.pone.0191154 |
| 4 | Jaakkola and Elo, 2016 | https://doi.org/10.1093/bib/bbv049 |
| 5 | Ma et al., 2019 | https://doi.org/10.1186/s12859-019-3146-1 |
| 6 | Maleki et al., 2019a | https://doi.org/10.1142/S0219720019400109 |
| 7 | Mathur et al., 2018 | https://doi.org/10.1186/s13040-018-0166-8 |
| 8 | Nguyen et al., 2019 | https://doi.org/10.1186/s13059-019-1790-4 |
| 9 | Rahmatallah et al., 2016 | https://doi.org/10.1093/bib/bbv069 |
| 10 | Tarca et al., 2013 | https://doi.org/10.1371/journal.pone.0079217 |
| 11 | Zyla et al., 2017a | https://doi.org/10.1007/978-3-319-60816-7_18 |
| 12 | Zyla et al., 2019 | https://doi.org/10.1093/bioinformatics/btz447 |
| 13 | Maciejewski (2014) | https://doi.org/10.1093/bib/bbt002 |

# STable 2 - Methods

| Test/Method name | DOI | Reference | \multicolumn{13}{c}{Study (Refer to STable 1)} |
|---|---|---|---|---|---|---|---|---|---|---|---|---|---|---|---|
| | | | 1 | 2 | 3 | 4 | 5 | 6 | 7 | 8 | 9 | 10 | 11 | 12 | Total |
| CAMERA | https://doi.org/10.1093/nar/gks461 | Wu et al. (2012) | | x | | x | x | x | | | x | | | | 5 |
| CePa ORA | https://doi.org/10.1186/1752-0509-6-56 | Gu et al. (2012) | x | | x | x | x | | | x | | | | | 5 |
| CePa GSA | https://doi.org/10.1093/bioinformatics/btt008 | Gu and Wang (2013) | x | | | | | | | x | | | | | 2 |
| CERNO | https://doi.org/10.1016/j.jneuroim.2007.12.007 | Yamaguchi et al. (2008) | | | | | | | | | | | | x | 1 |
| Clipper | https://doi.org/10.1093/nar/gks866 | Martini et al. (2013) | | x | | | | | | | | | | | 1 |
| DEGraph | https://doi.org/10.1214/11-AOAS528 | Jacob et al. (2012) | | x | | x | | | | | | | | | 2 |
| DESeq + Fisher's method | https://doi.org/10.1186/gb-2010-11-10-r106 | Anders and Huber (2010) | | | | | | | | | | x | | | 1 |
| eBayes + Fisher's method | https://doi.org/10.1007/0-387-29362-0_23 | Smyth (2005) | | | | | | | | | | x | | | 1 |
| edgeR + Fisher's method | https://doi.org/10.1093/bioinformatics/btp616 | Robinson et al. (2010) | | | | | | | | | | x | | | 1 |
| FRY | https://doi.org/10.1093/nar/gkv007 | Ritchie et al. (2015) | | | | | | x | | | | | | | 1 |
| GAGE | https://doi.org/10.1186/1471-2105-10-161 | Luo et al. (2009) | | | | | | x | | | | x | | | 2 |
| GeneSetTest/MRGSE | https://doi.org/10.1186/1471-2164-9-363 | Michaud et al. (2008) | | | | | | | | | | x | x | | 2 |
| GlobalTest | https://doi.org/10.1093/bioinformatics/btg382 | Goeman et al. (2004) | | x | | | x | | | | | x | x | | 4 |
| GSA | https://doi.org/10.1214/07-AOAS101 | Efron and Tibshirani (2007) | | x | | | | | | x | | x | | | 3 |
| GSEA-G (gene permutation) | https://doi.org/10.1073/pnas.0506580102 | Subramanian et al. (2005) | | | | | | x | x | | | x | | | 3 |
| GSEA-S (sample permutation) | https://doi.org/10.1073/pnas.0506580102 | Subramanian et al. (2005) | | x | | x | | x | x | x | | x | x | x | 8 |
| GSVA | https://doi.org/10.1186/1471-2105-14-7 | Hänzelmann et al. (2013) | | x | | | | x | | | | x | x | x | x | 6 |
| Kolmogorov-Smirnov | https://doi.org/10.1080/01621459.1951.10500769 | Massey Jr. (1951) | x | | | | | | | x | x | | | | 3 |
| N-Statistic | https://doi.org/10.1016/S0047-259X(03)00079-4 | Baringhaus and Franz (2004) | | | | | | | | | | x | | | 1 |
| NetGSA | https://doi.org/10.2202/1544-6115.1483 | Shojaie and George (2013) | | | | x | x | | | | | | | | 2 |
| ORA / Fisher's test (or variant) | - | - | x | x | | x | | x | | x | | x | x | x | 8 |
| PADOG | https://doi.org/10.1186/1471-2105-13-136 | Tarca et al. (2012) | | x | | | | x | | x | | x | x | x | 6 |
| PAGE | https://doi.org/10.1186/1471-2105-6-144 | Kim and Volsky (2005) | | | | | | | x | | | | | | 1 |
| PathNet | https://doi.org/10.1186/1751-0473-7-10 | Dutta et al. (2012) | x | | | | x | | | x | | | | | 3 |
| Pathway-Express | https://doi.org/10.1101/gr.6202607 | Drăghici et al. (2007) | | | | | | | x | | | | | | 1 |
| Pathifier | https://doi.org/10.1073/pnas.1219651110 | Drier et al. (2012) | | | | x | | | | | | | | | 1 |
| PLAGE | https://doi.org/10.1186/1471-2105-6-225 | Tomfohr et al. (2005) | | | | | | | x | | | x | x | x | 4 |
| PRS | https://doi.org/10.1089/cmb.2011.0182 | Al-Haj Ibrahim et al. (2012) | | | x | | x | | | | | | | | 2 |

| Test/Method name | DOI | Reference | | | | | | | | | | | |
|---|---|---|---|---|---|---|---|---|---|---|---|---|---|
| ROAST | https://doi.org/10.1093/bioinformatics/btq401 | Wu et al. (2010) | | x | | | x | | x | | | | 3 |
| ROMER | https://doi.org/10.1093/nar/gkv007 | Ritchie et al. (2015) | | | | | | | x | | | | 1 |
| SAFE-Wilcoxon rank sum | https://doi.org/10.1093/bioinformatics/bti260 | Barry et al. (2005) | | x | | | | x | | x | | | 3 |
| SAFE-Fisher's exact test | https://doi.org/10.1093/bioinformatics/bti260 | Barry et al. (2005) | | | | | | x | | | | | 1 |
| SAFE-Pearson's Chi-squared test | https://doi.org/10.1093/bioinformatics/bti260 | Barry et al. (2005) | | | | | | x | | | | | 1 |
| SAFE-average difference | https://doi.org/10.1093/bioinformatics/bti260 | Barry et al. (2005) | | | | | | x | | | | | 1 |
| SAM-GS | https://doi.org/10.1186/1471-2105-8-242 | Dinu et al. (2007) | | x | | | | | | x | | | 2 |
| seqGSEA | https://doi.org/10.1186/1471-2105-14-S5-S16 | Wang and Cairns (2013) | | | | | | | | x | | | 1 |
| SIGPATHWAYQ1 | https://doi.org/10.1073/pnas.0506577102 | Tian et al. (2005) | | | | | | x | | x | | | 2 |
| SIGPATHWAYQ2 | https://doi.org/10.1073/pnas.0506577102 | Tian et al. (2005) | | | | | | | | x | | | 1 |
| SPIA | https://doi.org/10.1093/bioinformatics/btn577 | Tarca et al. (2009) | x | | x | x | x | | x | | | | 5 |
| ssGSEA | https://doi.org/10.1038/nature08460 | Barbie et al. (2009) | | | | | x | | x | x | | | 3 |
| TAPPA | https://doi.org/10.1093/bioinformatics/btm460 | Gao et al. (2007) | | | x | | | | | | | | 1 |
| topologyGSA | https://doi.org/10.1186/1752-0509-4-121 | Massa et al. (2010) | | | x | x | | | | | | | 2 |
| Wilcoxon rank sum | https://doi.org/10.2307/3001968 | Wilcoxon (1945) | x | | | | | | x | | | | 2 |
| Wilcoxon GST | https://doi.org/10.1007/0-387-29362-0_23 | Smyth (2005) | | | | | | | | | x | | 1 |
| Z-Score | - | - | | | | | | | | x | | | 1 |
| | | | | | | | | | | | | | |
| **Total number of methods per study** | | | 7 | 10 | 7 | 6 | 9 | 13 | 8 | 10 | 11 | 16 | 5 | 9 |

## STable 3 - Sensitivity

| | | | Study (Refer to STable 1) | | | | | | |
|---|---|---|---|---|---|---|---|---|---|
| Test/Method name | DOI | Reference | #1 | #3 | #8 | #9 | #10 | #11 | #12 |
| CAMERA | https://doi.org/10.1093/nar/gks461 | Wu et al. (2012) | | | | | "--" | | |
| CePa ORA | https://doi.org/10.1186/1752-0509-6-56 | Gu et al. (2012) | "+-" | "+-" | | | | | |
| CePa GSA | https://doi.org/10.1093/bioinformatics/btt008 | Gu and Wang (2013) | "+-" | | | | | | |
| CERNO | https://doi.org/10.1016/j.jneuroim.2007.12.007 | Yamaguchi et al. (2008) | | | | | | | "+" |
| Clipper | https://doi.org/10.1093/nar/gks866 | Martini et al. (2013) | | "+-" | | | | | |
| DEGraph | https://doi.org/10.1214/11-AOAS528 | Jacob et al. (2012) | | "+-" | | | | | |
| DESeq + Fisher's method | https://doi.org/10.1186/gb-2010-11-10-r106 | Anders and Huber (2010) | | | | "+" | | | |

| Method | DOI | Reference | | | | | | |
|---|---|---|---|---|---|---|---|---|
| eBayes + Fisher's method | https://doi.org/10.1007/0-387-29362-0_23 | Smyth (2005) | | | | "+" | | |
| edgeR + Fisher's method | https://doi.org/10.1093/bioinformatics/btp616 | Robinson et al. (2010) | | | | "+" | | |
| FRY | https://doi.org/10.1093/nar/gkv007 | Ritchie et al. (2015) | | | | | | |
| GAGE | https://doi.org/10.1186/1471-2105-10-161 | Luo et al. (2009) | | | | "++" | | |
| GeneSetTest/MRGSE | https://doi.org/10.1186/1471-2164-9-363 | Michaud et al. (2008) | | | | "++" | | "+" |
| GlobalTest | https://doi.org/10.1093/bioinformatics/btg382 | Goeman et al. (2004) | | | | "++" | | "+" |
| GSA | https://doi.org/10.1214/07-AOAS101 | Efron and Tibshirani (2007) | | | "+" | "--" | | |
| GSEA-G (gene permutation) | https://doi.org/10.1073/pnas.0506580102 | Subramanian et al. (2005) | | | | "-" | | "+" |
| GSEA-S (sample permutation) | https://doi.org/10.1073/pnas.0506580102 | Subramanian et al. (2005) | | "--" | | "+" | | |
| GSVA | https://doi.org/10.1073/pnas.0506580102 | Subramanian et al. (2005) | | | "-" | "--" | | "--" |
| Kolmogorov-Smirnov | https://doi.org/10.1080/01621459.1951.10500769 | Massey Jr. (1951) | "++" | | "++" | "+" | | |
| N-Statistic | https://doi.org/10.1016/S0047-259X(03)00079-4 | Baringhaus and Franz (2004) | | | | "++" | | |
| NetGSA | https://doi.org/10.2202/1544-6115.1483 | Shojaie and George (2013) | | | | | | |
| ORA / Fisher's test (or variant) | - | - | "+-" | | "--" | "+" | "-" | "+" |
| PADOG | https://doi.org/10.1186/1471-2105-13-136 | Tarca et al. (2012) | | | "-" | "-" | "-" | "++" |
| PAGE | https://doi.org/10.1186/1471-2105-6-144 | Kim et al. (2005) | | | | | | |
| PathNet | https://doi.org/10.1186/1751-0473-7-10 | Dutta et al. (2012) | "+-" | | | | | |
| Pathway-Express | https://doi.org/10.1101/gr.6202607 | Drăghici et al. (2007) | | | | | | |
| Pathifier | https://doi.org/10.1073/pnas.1219651110 | Drier et al. (2012) | | | | | | |
| PLAGE | https://doi.org/10.1186/1471-2105-6-225 | Tomfohr et al. (2005) | | | | "++" | | "+" |
| PRS | https://doi.org/10.1089/cmb.2011.0182 | Al-Haj Ibrahim et al. (2012) | | "+-" | | | | |
| ROAST | https://doi.org/10.1093/bioinformatics/btq401 | Wu et al. (2010) | | | "+" | | | |
| ROMER | https://doi.org/10.1093/nar/gkv007 | Ritchie et al. (2015) | | | "-" | | | |
| SAFE | https://doi.org/10.1093/bioinformatics/bti260 | Barry et al. (2005) | | | | "-" | | |
| SAM-GS | https://doi.org/10.1186/1471-2105-8-242 | Dinu et al. (2007) | | | "++" | | | |
| seqGSEA | https://doi.org/10.1186/1471-2105-14-S5-S16 | Wang and Cairns (2013) | | | "+" | | | |
| SIGPATHWAYQ1 | https://doi.org/10.1073/pnas.0506577102 | Tian et al. (2005) | | | | "-" | | |
| SIGPATHWAYQ2 | https://doi.org/10.1073/pnas.0506577102 | Tian et al. (2005) | | | | "+" | | |
| SPIA | https://doi.org/10.1093/bioinformatics/btn577 | Tarca et al. (2009) | "+-" | "+-" | "+" | | | |
| ssGSEA | https://doi.org/10.1038/nature08460 | Barbie et al. (2009) | | | | "+" | "+" | |
| TAPPA | https://doi.org/10.1093/bioinformatics/btm460 | Gao et al. (2007) | | "+-" | | | | |
| topologyGSA | https://doi.org/10.1186/1752-0509-4-121 | Massa et al. (2010) | | "+-" | "--" | | | |

| Test/Method name | DOI | Reference | | | | | | |
|---|---|---|---|---|---|---|---|---|
| Wilcoxon rank sum | https://doi.org/10.2307/3001968 | Wilcoxon (1945) | "++" | | "++" | | | |
| Wilcoxon GST | https://doi.org/10.1007/0-387-29362-0_23 | Smyth (2005) | | | | | | |
| Z-Score | - | - | | | | "-" | | |

# STable 4 – Specificity

| | | | Study (Refer to STable 1) | | | | | | | |
|---|---|---|---|---|---|---|---|---|---|---|
| Test/Method name | DOI | Reference | #1 | #3 | #4 | #5 | #8 | #10 | #11 | #12 |
| CAMERA | https://doi.org/10.1093/nar/gks461 | Wu and Smyth (2012) | | | | "++" | | "+" | | |
| CePa ORA | https://doi.org/10.1186/1752-0509-6-56 | Gu et al. (2012) | "+-" | "+" | "-" | | | | | |
| CePa GSA | https://doi.org/10.1093/bioinformatics/btt008 | Gu and Wang (2013) | "+-" | | | | | | | |
| CERNO | https://doi.org/10.1016/j.jneuroim.2007.12.007 | Yamaguchi et al. (2008) | | | | | | | "+" | |
| Clipper | https://doi.org/10.1093/nar/gks866 | Martini et al. (2013) | | "-" | | | | | | |
| DEGraph | https://doi.org/10.1214/11-AOAS528 | Jacob et al. (2012) | | "+" | | "+" | | | | |
| DESeq + Fisher's method | https://doi.org/10.1186/gb-2010-11-10-r106 | Anders and Huber (2010) | | | | | | | | |
| eBayes + Fisher's method | https://doi.org/10.1007/0-387-29362-0_23 | Smyth (2005) | | | | | | | | |
| edgeR + Fisher's method | https://doi.org/10.1093/bioinformatics/btp616 | Robinson et al. (2010) | | | | | | | | |
| FRY | https://doi.org/10.1093/nar/gkv007 | Ritchie et al. (2015) | | | | | | | | |
| GAGE | https://doi.org/10.1186/1471-2105-10-161 | Luo et al. (2009) | | | | | | "--" | | |
| GeneSetTest/MRGSE | https://doi.org/10.1186/1471-2164-9-363 | Michaud et al. (2008) | | | | | | "-" | | "--" |
| GlobalTest | https://doi.org/10.1093/bioinformatics/btg382 | Goeman et al. (2004) | | | | | | "+" | | "++" |
| GSA | https://doi.org/10.1214/07-AOAS101 | Efron and Tibshirani (2007) | | | | | "-" | "+" | | |
| GSEA-G (gene permutation) | https://doi.org/10.1073/pnas.0506580102 | Subramanian et al. (2005) | | | | | | "-" | | "+-" |
| GSEA-S (sample permutation) | https://doi.org/10.1073/pnas.0506580102 | Subramanian et al. (2005) | | | "-" | | "--" | "+" | "+" | |
| GSVA | https://doi.org/10.1073/pnas.0506580102 | Subramanian et al. (2005) | | | | | | "+" | "+" | "-" |
| Kolmogorov-Smirnov | https://doi.org/10.1080/01621459.1951.10500769 | Massey Jr. (1951) | "+-" | | | | "--" | | | |
| N-Statistic | https://doi.org/10.1016/S0047-259X(03)00079-4 | Baringhaus and Franz (2004) | | | | | | | | |
| NetGSA | https://doi.org/10.2202/1544-6115.1483 | Shojaie and George (2013) | | | "-" | "+-" | | | | |
| ORA / Fisher's test (or variant) | - | - | "+-" | | "-" | | "-" | "+" | "-" | "+" |
| PADOG | https://doi.org/10.1186/1471-2105-13-136 | Tarca et al. (2012) | | | | "+" | | "+" | "-" | "+" |
| PAGE | https://doi.org/10.1186/1471-2105-6-144 | Kim et al. (2005) | | | | | | | | |
| PathNet | https://doi.org/10.1186/1751-0473-7-10 | Dutta et al. (2012) | "+-" | | | "--" | | | | |

| Test/Method name | DOI | Reference | | | | | | | |
|---|---|---|---|---|---|---|---|---|---|
| Pathway-Express | https://doi.org/10.1101/gr.6202607 | Drăghici et al. (2007) | | | | "+-" | | | |
| Pathifier | https://doi.org/10.1073/pnas.1219651110 | Drier et al. (2012) | | | "--" | | | | |
| PLAGE | https://doi.org/10.1186/1471-2105-6-225 | Tomfohr et al. (2005) | | | | | "+" | "+" | "-" |
| PRS | https://doi.org/10.1089/cmb.2011.0182 | Al-Haj Ibrahim et al. (2012) | | "+" | | | | | |
| ROAST | https://doi.org/10.1093/bioinformatics/btq401 | Wu et al. (2010) | | | | | | | |
| ROMER | https://doi.org/10.1093/nar/gkv007 | Ritchie et al. (2015) | | | | | | | |
| SAFE-Wilcoxon rank sum | https://doi.org/10.1093/bioinformatics/bti260 | Barry et al. (2005) | | | | | "+" | | |
| SAM-GS | https://doi.org/10.1186/1471-2105-8-242 | Dinu et al. (2007) | | | | | | | |
| seqGSEA | https://doi.org/10.1186/1471-2105-14-S5-S16 | Wang and Cairns (2013) | | | | | | | |
| SIGPATHWAYQ1 | https://doi.org/10.1073/pnas.0506577102 | Tian et al. (2005) | | | | | "-" | | |
| SIGPATHWAYQ2 | https://doi.org/10.1073/pnas.0506577102 | Tian et al. (2005) | | | | | "+" | | |
| SPIA | https://doi.org/10.1093/bioinformatics/btn577 | Tarca et al. (2009) | "+-" | "+" | "+" | | "+" | | |
| ssGSEA | https://doi.org/10.1038/nature08460 | Barbie et al. (2009) | | | | | "+" | | |
| TAPPA | https://doi.org/10.1093/bioinformatics/btm460 | Gao et al. (2007) | | "+" | | | | | |
| topologyGSA | https://doi.org/10.1186/1752-0509-4-121 | Massa et al. (2010) | | "-" | | "+-" | | | |
| Wilcoxon rank sum | https://doi.org/10.2307/3001968 | Wilcoxon (1945) | "+-" | | | | "+" | | |
| Wilcoxon GST | https://doi.org/10.1007/0-387-29362-0_23 | Smyth (2005) | | | | | | | |
| Z-Score | - | - | | | | | "+" | | |

## STable 5 – Prioritization

| | | | Study (Refer to STable 1) | | | | |
|---|---|---|---|---|---|---|---|
| Test/Method name | DOI | Reference | #1 | #3 | #8 | #10 | #12 |
| CAMERA | https://doi.org/10.1093/nar/gks461 | Wu et al. (2012) | | | | "-" | |
| CePa ORA | https://doi.org/10.1186/1752-0509-6-56 | Gu et al. (2012) | "-" | "+-" | "+" | | |
| CePa GSA | https://doi.org/10.1093/bioinformatics/btt008 | Gu and Wang (2013) | "-" | | "+" | | |
| CERNO | https://doi.org/10.1016/j.jneuroim.2007.12.007 | Yamaguchi et al. (2008) | | | | | "+" |
| Clipper | https://doi.org/10.1093/nar/gks866 | Martini et al. (2013) | | "++" | | | |
| DEGraph | https://doi.org/10.1214/11-AOAS528 | Jacob et al. (2012) | | "++" | | | |
| DESeq + Fisher's method | https://doi.org/10.1186/gb-2010-11-10-r106 | Anders and Huber (2010) | | | | | |
| eBayes + Fisher's method | https://doi.org/10.1007/0-387-29362-0_23 | Smyth (2005) | | | | | |
| edgeR + Fisher's method | https://doi.org/10.1093/bioinformatics/btp616 | Robinson et al. (2010) | | | | | |

| Method | DOI | Reference | | | | | |
|---|---|---|---|---|---|---|---|
| FRY | https://doi.org/10.1093/nar/gkv007 | Ritchie et al. (2015) | | | | | |
| GAGE | https://doi.org/10.1186/1471-2105-10-161 | Luo et al. (2009) | | | | "-" | |
| GeneSetTest/MRGSE | https://doi.org/10.1186/1471-2164-9-363 | Michaud et al. (2008) | | | | "+" | "+" |
| GlobalTest | https://doi.org/10.1093/bioinformatics/btg382 | Goeman et al. (2004) | | | | "+" | "-" |
| GSA | https://doi.org/10.1214/07-AOAS101 | Efron and Tibshirani (2007) | | | "+" | "+" | |
| GSEA-G (gene permutation) | https://doi.org/10.1073/pnas.0506580102 | Subramanian et al. (2005) | | | | "-" | |
| GSEA-S (sample permutation) | https://doi.org/10.1073/pnas.0506580102 | Subramanian et al. (2005) | | | "+" | "-" | "+" |
| GSVA | https://doi.org/10.1073/pnas.0506580102 | Subramanian et al. (2005) | | | | "--" | "--" |
| Kolmogorov-Smirnov | https://doi.org/10.1080/01621459.1951.10500769 | Massey Jr. (1951) | "+" | | "--" | | |
| N-Statistic | https://doi.org/10.1016/S0047-259X(03)00079-4 | Baringhaus and Franz (2004) | | | | | |
| NetGSA | https://doi.org/10.2202/1544-6115.1483 | Shojaie and George (2013) | | | | | |
| ORA / Fisher's test (or variant) | - | - | "-" | | "-" | "++" | "+-" |
| PADOG | https://doi.org/10.1186/1471-2105-13-136 | Tarca et al. (2012) | | | "++" | "++" | "++" |
| PAGE | https://doi.org/10.1186/1471-2105-6-144 | Kim et al. (2005) | | | | | |
| PathNet | https://doi.org/10.1186/1751-0473-7-10 | Dutta et al. (2012) | "++" | | "+" | | |
| Pathway-Express | https://doi.org/10.1101/gr.6202607 | Drăghici et al. (2007) | | | | | |
| Pathifier | https://doi.org/10.1073/pnas.1219651110 | Drier et al. (2012) | | | | | |
| PLAGE | https://doi.org/10.1186/1471-2105-6-225 | Tomfohr et al. (2005) | | | | "+" | "+" |
| PRS | https://doi.org/10.1089/cmb.2011.0182 | Al-Haj Ibrahim et al. (2012) | | "+-" | | | |
| ROAST | https://doi.org/10.1093/bioinformatics/btq401 | Wu et al. (2010) | | | | | |
| ROMER | https://doi.org/10.1093/nar/gkv007 | Ritchie et al. (2015) | | | | | |
| SAFE-Wilcoxon rank sum | https://doi.org/10.1093/bioinformatics/bti260 | Barry et al. (2005) | | | | "+" | |
| SAM-GS | https://doi.org/10.1186/1471-2105-8-242 | Dinu et al. (2007) | | | | | |
| seqGSEA | https://doi.org/10.1186/1471-2105-14-S5-S16 | Wang and Cairns (2013) | | | | | |
| SIGPATHWAYQ1 | https://doi.org/10.1073/pnas.0506577102 | Tian et al. (2005) | | | | "--" | |
| SIGPATHWAYQ2 | https://doi.org/10.1073/pnas.0506577102 | Tian et al. (2005) | | | | "-" | |
| SPIA | https://doi.org/10.1093/bioinformatics/btn577 | Tarca et al. (2009) | "-" | "+-" | "+" | | |
| ssGSEA | https://doi.org/10.1038/nature08460 | Barbie et al. (2009) | | | | "-" | |
| TAPPA | https://doi.org/10.1093/bioinformatics/btm460 | Gao et al. (2007) | | "+-" | | | |
| topologyGSA | https://doi.org/10.1186/1752-0509-4-121 | Massa et al. (2010) | | "++" | | | |
| Wilcoxon rank sum | https://doi.org/10.2307/3001968 | Wilcoxon (1945) | "+" | | "-" | | |
| Wilcoxon GST | https://doi.org/10.1007/0-387-29362-0_23 | Smyth (2005) | | | | | |
| Z-Score | - | - | | | | "-" | |

# STable 6 – Sample size robustness

| | | | Study (Refer to STable 1) | | | |
|---|---|---|---|---|---|---|
| Test/Method name | DOI | Reference | #3 | #9 | #10 | #12 |
| CAMERA | https://doi.org/10.1093/nar/gks461 | Wu et al. (2012) | | | "--" | |
| CePa ORA | https://doi.org/10.1186/1752-0509-6-56 | Gu et al. (2012) | "++" | | | |
| CePa GSA | https://doi.org/10.1093/bioinformatics/btt008 | Gu and Wang (2013) | | | | |
| CERNO | https://doi.org/10.1016/j.jneuroim.2007.12.007 | Yamaguchi et al. (2008) | | | | "+" |
| Clipper | https://doi.org/10.1093/nar/gks866 | Martini et al. (2013) | "--" | | | |
| DEGraph | https://doi.org/10.1214/11-AOAS528 | Jacob et al. (2012) | "--" | | | |
| DESeq + Fisher's method | https://doi.org/10.1186/gb-2010-11-10-r106 | Anders and Huber (2010) | | "+" | | |
| eBayes + Fisher's method | https://doi.org/10.1007/0-387-29362-0_23 | Smyth (2005) | | "+" | | |
| edgeR + Fisher's method | https://doi.org/10.1093/bioinformatics/btp616 | Robinson et al. (2010) | | "+" | | |
| FRY | https://doi.org/10.1093/nar/gkv007 | Ritchie et al. (2015) | | | | |
| GAGE | https://doi.org/10.1186/1471-2105-10-161 | Luo et al. (2009) | | | "+" | |
| GeneSetTest/MRGSE | https://doi.org/10.1186/1471-2164-9-363 | Michaud et al. (2008) | | | "++" | "+" |
| GlobalTest | https://doi.org/10.1093/bioinformatics/btg382 | Goeman et al. (2004) | | | "++" | "-" |
| GSA | https://doi.org/10.1214/07-AOAS101 | Efron and Tibshirani (2007) | | | "-" | |
| GSEA-G (gene permutation) | https://doi.org/10.1073/pnas.0506580102 | Subramanian et al. (2005) | | | "+" | "+" |
| GSEA-S (sample permutation) | https://doi.org/10.1073/pnas.0506580102 | Subramanian et al. (2005) | | | "-" | |
| GSVA | https://doi.org/10.1073/pnas.0506580102 | Subramanian et al. (2005) | | "-" | "-" | "+" |
| Kolmogorov-Smirnov | https://doi.org/10.1080/01621459.1951.10500769 | Massey Jr. (1951) | | "+" | | |
| N-Statistic | https://doi.org/10.1016/S0047-259X(03)00079-4 | Baringhaus and Franz (2004) | | "++" | | |
| NetGSA | https://doi.org/10.2202/1544-6115.1483 | Shojaie and George (2013) | | | | |
| ORA / Fisher's test (or variant) | - | - | | | "+" | "+" |
| PADOG | https://doi.org/10.1186/1471-2105-13-136 | Tarca et al. (2012) | | | "+" | "+" |
| PAGE | https://doi.org/10.1186/1471-2105-6-144 | Kim et al. (2005) | | | | |
| PathNet | https://doi.org/10.1186/1751-0473-7-10 | Dutta et al. (2012) | | | | |
| Pathway-Express | https://doi.org/10.1101/gr.6202607 | Drăghici et al. (2007) | | | | |
| Pathifier | https://doi.org/10.1073/pnas.1219651110 | Drier et al. (2012) | | | | |
| PLAGE | https://doi.org/10.1186/1471-2105-6-225 | Tomfohr et al. (2005) | | | "+" | "-" |
| PRS | https://doi.org/10.1089/cmb.2011.0182 | Al-Haj Ibrahim et al. (2012) | "++" | | | |
| ROAST | https://doi.org/10.1093/bioinformatics/btq401 | Wu et al. (2010) | | "+" | | |
| ROMER | https://doi.org/10.1093/nar/gkv007 | Ritchie et al. (2015) | | "-" | | |
| SAFE-Wilcoxon rank sum | https://doi.org/10.1093/bioinformatics/bti260 | Barry et al. (2005) | | | "-" | |

| Test/Method name | DOI | Reference | | | | |
|---|---|---|---|---|---|---|
| SAM-GS | https://doi.org/10.1186/1471-2105-8-242 | Dinu et al. (2007) | | "++" | | |
| seqGSEA | https://doi.org/10.1186/1471-2105-14-S5-S16 | Wang and Cairns (2013) | | "+" | | |
| SIGPATHWAYQ1 | https://doi.org/10.1073/pnas.0506577102 | Tian et al. (2005) | | | "+" | |
| SIGPATHWAYQ2 | https://doi.org/10.1073/pnas.0506577102 | Tian et al. (2005) | | | "-" | |
| SPIA | https://doi.org/10.1093/bioinformatics/btn577 | Tarca et al. (2009) | "++" | | | |
| ssGSEA | https://doi.org/10.1038/nature08460 | Barbie et al. (2009) | | "+" | "-" | |
| TAPPA | https://doi.org/10.1093/bioinformatics/btm460 | Gao et al. (2007) | "-" | | | |
| topologyGSA | https://doi.org/10.1186/1752-0509-4-121 | Massa et al. (2010) | "--" | | | |
| Wilcoxon rank sum | https://doi.org/10.2307/3001968 | Wilcoxon (1945) | | | | |
| Wilcoxon GST | https://doi.org/10.1007/0-387-29362-0_23 | Smyth (2005) | | | | "+" |
| Z-Score | - | - | | | "-" | |

## STable 7 – Gene size robustness

| | | | Study (Refer to STable 1) | | | |
|---|---|---|---|---|---|---|
| Test/Method name | DOI | Reference | #1 | #2 | #10 | #12 |
| CAMERA | https://doi.org/10.1093/nar/gks461 | Wu et al. (2012) | | "+/-" | "-" | |
| CePa ORA | https://doi.org/10.1186/1752-0509-6-56 | Gu et al. (2012) | "+/-" | | | |
| CePa GSA | https://doi.org/10.1093/bioinformatics/btt008 | Gu and Wang (2013) | "+" | | | |
| CERNO | https://doi.org/10.1016/j.jneuroim.2007.12.007 | Yamaguchi et al. (2008) | | | | "+" |
| Clipper | https://doi.org/10.1093/nar/gks866 | Martini et al. (2013) | | | | |
| DEGraph | https://doi.org/10.1214/11-AOAS528 | Jacob et al. (2012) | | | | |
| DESeq + Fisher's method | https://doi.org/10.1186/gb-2010-11-10-r106 | Anders and Huber (2010) | | | | |
| eBayes + Fisher's method | https://doi.org/10.1007/0-387-29362-0_23 | Smyth (2005) | | | | |
| edgeR + Fisher's method | https://doi.org/10.1093/bioinformatics/btp616 | Robinson et al. (2010) | | | | |
| FRY | https://doi.org/10.1093/nar/gkv007 | Ritchie et al. (2015) | | | | |
| GAGE | https://doi.org/10.1186/1471-2105-10-161 | Luo et al. (2009) | | | "+" | |
| GeneSetTest/MRGSE | https://doi.org/10.1186/1471-2164-9-363 | Michaud et al. (2008) | | | "++" | "+" |
| GlobalTest | https://doi.org/10.1093/bioinformatics/btg382 | Goeman et al. (2004) | | "--" | "+/-" | "-" |
| GSA | https://doi.org/10.1214/07-AOAS101 | Efron and Tibshirani (2007) | | "+/-" | "-" | |
| GSEA-G (gene permutation) | https://doi.org/10.1073/pnas.0506580102 | Subramanian et al. (2005) | | | "++" | "+" |
| GSEA-S (sample permutation) | https://doi.org/10.1073/pnas.0506580102 | Subramanian et al. (2005) | | "+/-" | "-" | |
| GSVA | https://doi.org/10.1073/pnas.0506580102 | Subramanian et al. (2005) | | "+/-" | "-" | "+" |
| Kolmogorov-Smirnov | https://doi.org/10.1080/01621459.1951.10500769 | Massey Jr. (1951) | "+" | | | |
| N-Statistic | https://doi.org/10.1016/S0047-259X(03)00079-4 | Baringhaus and Franz (2004) | | | | |

| Test/Method name | DOI | Reference | | | | |
|---|---|---|---|---|---|---|
| NetGSA | https://doi.org/10.2202/1544-6115.1483 | Shojaie and George (2013) | | | | |
| ORA / Fisher's test (or variant) | - | - | "+/-" | "+/-" | "+/-" | "+" |
| PADOG | https://doi.org/10.1186/1471-2105-13-136 | Tarca et al. (2012) | | "+/-" | "++" | "+" |
| PAGE | https://doi.org/10.1186/1471-2105-6-144 | Kim et al. (2005) | | | | |
| PathNet | https://doi.org/10.1186/1751-0473-7-10 | Dutta et al. (2012) | "+" | | | |
| Pathway-Express | https://doi.org/10.1101/gr.6202607 | Drăghici et al. (2007) | | | | |
| Pathifier | https://doi.org/10.1073/pnas.1219651110 | Drier et al. (2012) | | | | |
| PLAGE | https://doi.org/10.1186/1471-2105-6-225 | Tomfohr et al. (2005) | | | "+" | "-" |
| PRS | https://doi.org/10.1089/cmb.2011.0182 | Al-Haj Ibrahim et al. (2012) | | | | |
| ROAST | https://doi.org/10.1093/bioinformatics/btq401 | Wu et al. (2010) | | "+/-" | | |
| ROMER | https://doi.org/10.1093/nar/gkv007 | Ritchie et al. (2015) | | | | |
| SAFE-Wilcoxon rank sum | https://doi.org/10.1093/bioinformatics/bti260 | Barry et al. (2005) | | "+/-" | "+/-" | |
| SAM-GS | https://doi.org/10.1186/1471-2105-8-242 | Dinu et al. (2007) | | "--" | | |
| seqGSEA | https://doi.org/10.1186/1471-2105-14-S5-S16 | Wang and Cairns (2013) | | | | |
| SIGPATHWAYQ1 | https://doi.org/10.1073/pnas.0506577102 | Tian et al. (2005) | | | "+" | |
| SIGPATHWAYQ2 | https://doi.org/10.1073/pnas.0506577102 | Tian et al. (2005) | | | "+/-" | |
| SPIA | https://doi.org/10.1093/bioinformatics/btn577 | Tarca et al. (2009) | "+/-" | | | |
| ssGSEA | https://doi.org/10.1038/nature08460 | Barbie et al. (2009) | | | "-" | |
| TAPPA | https://doi.org/10.1093/bioinformatics/btm460 | Gao et al. (2007) | | | | |
| topologyGSA | https://doi.org/10.1186/1752-0509-4-121 | Massa et al. (2010) | | | | |
| Wilcoxon rank sum | https://doi.org/10.2307/3001968 | Wilcoxon (1945) | "+" | | | |
| Wilcoxon GST | https://doi.org/10.1007/0-387-29362-0_23 | Smyth (2005) | | | | "+" |
| Z-Score | - | - | | | "-" | |

## STable 8 – Phenotype relevance

| | | | Study (Refer to STable 1) | |
|---|---|---|---|---|
| Test/Method name | DOI | Reference | #2 | #6 |
| CAMERA | https://doi.org/10.1093/nar/gks461 | Wu et al. (2012) | "+" | "-" |
| CePa ORA | https://doi.org/10.1186/1752-0509-6-56 | Gu et al. (2012) | | |
| CePa GSA | https://doi.org/10.1093/bioinformatics/btt008 | Gu and Wang (2013) | | |
| CERNO | https://doi.org/10.1016/j.jneuroim.2007.12.007 | Yamaguchi et al. (2008) | | |
| Clipper | https://doi.org/10.1093/nar/gks866 | Martini et al. (2013) | | |

| Method | DOI | Reference | | |
|---|---|---|---|---|
| DEGraph | https://doi.org/10.1214/11-AOAS528 | Jacob et al. (2012) | | |
| DESeq + Fisher's method | https://doi.org/10.1186/gb-2010-11-10-r106 | Anders and Huber (2010) | | |
| eBayes + Fisher's method | https://doi.org/10.1007/0-387-29362-0_23 | Smyth (2005) | | |
| edgeR + Fisher's method | https://doi.org/10.1093/bioinformatics/btp616 | Robinson et al. (2010) | | |
| FRY | https://doi.org/10.1093/nar/gkv007 | Ritchie et al. (2015) | | "-" |
| GAGE | https://doi.org/10.1186/1471-2105-10-161 | Luo et al. (2009) | | "+" |
| GeneSetTest/MRGSE | https://doi.org/10.1186/1471-2164-9-363 | Michaud et al. (2008) | | |
| GlobalTest | https://doi.org/10.1093/bioinformatics/btg382 | Goeman et al. (2004) | "-" | "-" |
| GSA | https://doi.org/10.1214/07-AOAS101 | Efron and Tibshirani (2007) | "+" | |
| GSEA-G (gene permutation) | https://doi.org/10.1073/pnas.0506580102 | Subramanian et al. (2005) | | "-" |
| GSEA-S (sample permutation) | https://doi.org/10.1073/pnas.0506580102 | Subramanian et al. (2005) | "+" | "+" |
| GSVA | https://doi.org/10.1073/pnas.0506580102 | Subramanian et al. (2005) | "+" | "-" |
| Kolmogorov-Smirnov | https://doi.org/10.1080/01621459.1951.10500769 | Massey Jr. (1951) | | |
| N-Statistic | https://doi.org/10.1016/S0047-259X(03)00079-4 | Baringhaus and Franz (2004) | | |
| NetGSA | https://doi.org/10.2202/1544-6115.1483 | Shojaie and George (2013) | | |
| ORA / Fisher's test (or variant) | - | - | "++" | "+" |
| PADOG | https://doi.org/10.1186/1471-2105-13-136 | Tarca et al. (2012) | "++" | "-" |
| PAGE | https://doi.org/10.1186/1471-2105-6-144 | Kim et al. (2005) | | "+" |
| PathNet | https://doi.org/10.1186/1751-0473-7-10 | Dutta et al. (2012) | | |
| Pathway-Express | https://doi.org/10.1101/gr.6202607 | Drăghici et al. (2007) | | |
| Pathifier | https://doi.org/10.1073/pnas.1219651110 | Drier et al. (2012) | | |
| PLAGE | https://doi.org/10.1186/1471-2105-6-225 | Tomfohr et al. (2005) | | "-" |
| PRS | https://doi.org/10.1089/cmb.2011.0182 | Al-Haj Ibrahim et al. (2012) | | |
| ROAST | https://doi.org/10.1093/bioinformatics/btq401 | Wu et al. (2010) | "-" | "-" |
| ROMER | https://doi.org/10.1093/nar/gkv007 | Ritchie et al. (2015) | | |
| SAFE-Wilcoxon rank sum | https://doi.org/10.1093/bioinformatics/bti260 | Barry et al. (2005) | "+" | |
| SAM-GS | https://doi.org/10.1186/1471-2105-8-242 | Dinu et al. (2007) | "-" | |
| seqGSEA | https://doi.org/10.1186/1471-2105-14-S5-S16 | Wang and Cairns (2013) | | |
| SIGPATHWAYQ1 | https://doi.org/10.1073/pnas.0506577102 | Tian et al. (2005) | | |
| SIGPATHWAYQ2 | https://doi.org/10.1073/pnas.0506577102 | Tian et al. (2005) | | |
| SPIA | https://doi.org/10.1093/bioinformatics/btn577 | Tarca et al. (2009) | | |
| ssGSEA | https://doi.org/10.1038/nature08460 | Barbie et al. (2009) | | "-" |
| TAPPA | https://doi.org/10.1093/bioinformatics/btm460 | Gao et al. (2007) | | |
| topologyGSA | https://doi.org/10.1186/1752-0509-4-121 | Massa et al. (2010) | | |

| Test/Method name | DOI | Reference | | |
|---|---|---|---|---|
| Wilcoxon rank sum | https://doi.org/10.2307/3001968 | Wilcoxon (1945) | | |
| Wilcoxon GST | https://doi.org/10.1007/0-387-29362-0_23 | Smyth (2005) | | |
| Z-Score | - | - | | |

# STable 9 – Accuracy

| | | | Study (Refer to STable 1) | |
|---|---|---|---|---|
| Test/Method name | DOI | Reference | #1 | #8 |
| CAMERA | https://doi.org/10.1093/nar/gks461 | Wu et al. (2012) | | |
| CePa ORA | https://doi.org/10.1186/1752-0509-6-56 | Gu et al. (2012) | "+-" | |
| CePa GSA | https://doi.org/10.1093/bioinformatics/btt008 | Gu and Wang (2013) | "+-" | |
| CERNO | https://doi.org/10.1016/j.jneuroim.2007.12.007 | Yamaguchi et al. (2008) | | |
| Clipper | https://doi.org/10.1093/nar/gks866 | Martini et al. (2013) | | |
| DEGraph | https://doi.org/10.1214/11-AOAS528 | Jacob et al. (2012) | | |
| DESeq + Fisher's method | https://doi.org/10.1186/gb-2010-11-10-r106 | Anders and Huber (2010) | | |
| eBayes + Fisher's method | https://doi.org/10.1007/0-387-29362-0_23 | Smyth (2005) | | |
| edgeR + Fisher's method | https://doi.org/10.1093/bioinformatics/btp616 | Robinson et al. (2010) | | |
| FRY | https://doi.org/10.1093/nar/gkv007 | Ritchie et al. (2015) | | |
| GAGE | https://doi.org/10.1186/1471-2105-10-161 | Luo et al. (2009) | | |
| GeneSetTest/MRGSE | https://doi.org/10.1186/1471-2164-9-363 | Michaud et al. (2008) | | |
| GlobalTest | https://doi.org/10.1093/bioinformatics/btg382 | Goeman et al. (2004) | | |
| GSA | https://doi.org/10.1214/07-AOAS101 | Efron and Tibshirani (2007) | | "+" |
| GSEA-G (gene permutation) | https://doi.org/10.1073/pnas.0506580102 | Subramanian et al. (2005) | | |
| GSEA-S (sample permutation) | https://doi.org/10.1073/pnas.0506580102 | Subramanian et al. (2005) | | "-" |
| GSVA | https://doi.org/10.1073/pnas.0506580102 | Subramanian et al. (2005) | | |
| Kolmogorov-Smirnov | https://doi.org/10.1080/01621459.1951.10500769 | Massey Jr. (1951) | "+-" | "-" |
| N-Statistic | https://doi.org/10.1016/S0047-259X(03)00079-4 | Baringhaus and Franz (2004) | | |
| NetGSA | https://doi.org/10.2202/1544-6115.1483 | Shojaie and George (2013) | | |
| ORA / Fisher's test (or variant) | - | - | "+-" | "-" |
| PADOG | https://doi.org/10.1186/1471-2105-13-136 | Tarca et al. (2012) | | "++" |
| PAGE | https://doi.org/10.1186/1471-2105-6-144 | Kim et al. (2005) | | |
| PathNet | https://doi.org/10.1186/1751-0473-7-10 | Dutta et al. (2012) | "+-" | |
| Pathway-Express | https://doi.org/10.1101/gr.6202607 | Drăghici et al. (2007) | | |
| Pathifier | https://doi.org/10.1073/pnas.1219651110 | Drier et al. (2012) | | |
| PLAGE | https://doi.org/10.1186/1471-2105-6-225 | Tomfohr et al. (2005) | | |
| PRS | https://doi.org/10.1089/cmb.2011.0182 | Al-Haj Ibrahim et al. (2012) | | |
| ROAST | https://doi.org/10.1093/bioinformatics/btq401 | Wu et al. (2010) | | |
| ROMER | https://doi.org/10.1093/nar/gkv007 | Ritchie et al. (2015) | | |
| SAFE-Wilcoxon rank sum | https://doi.org/10.1093/bioinformatics/bti260 | Barry et al. (2005) | | |
| SAM-GS | https://doi.org/10.1186/1471-2105-8-242 | Dinu et al. (2007) | | |
| seqGSEA | https://doi.org/10.1186/1471-2105-14-S5-S16 | Wang and Cairns (2013) | | |
| SIGPATHWAYQ1 | https://doi.org/10.1073/pnas.0506577102 | Tian et al. (2005) | | |
| SIGPATHWAYQ2 | https://doi.org/10.1073/pnas.0506577102 | Tian et al. (2005) | | |
| SPIA | https://doi.org/10.1093/bioinformatics/btn577 | Tarca et al. (2009) | "+-" | "+" |
| ssGSEA | https://doi.org/10.1038/nature08460 | Barbie et al. (2009) | | |
| TAPPA | https://doi.org/10.1093/bioinformatics/btm460 | Gao et al. (2007) | | |
| topologyGSA | https://doi.org/10.1186/1752-0509-4-121 | Massa et al. (2010) | | |
| Wilcoxon rank sum | https://doi.org/10.2307/3001968 | Wilcoxon (1945) | "+-" | "-" |
| Wilcoxon GST | https://doi.org/10.1007/0-387-29362-0_23 | Smyth (2005) | | |
| Z-Score | - | - | | |

## STable 10 – Additional measures

| Study (Refer to STable 1) | Type I errors | Proportion of significantly enriched gene sets | Reproducibility across datasets | Runtime | Power |
|---|---|---|---|---|---|
| 1 | | x | | | |
| 2 | x | x | | x | x |
| 3 | x | x | | x | |
| 5 | x | | | | |
| 6 | | | x | | |
| 7 | x | | | x | x |
| 9 | | x | | | |
| 11 | | | x | | |
| 12 | | | x | | |
| 13 | | | | | x |